\documentclass[]{aa}
\usepackage{txfonts,graphicx,natbib,subfigure}
\bibpunct{(}{)}{;}{a}{}{,} 
\begin{document}
\title{Enhanced phase mixing of Alfv\'en waves propagating in stratified and divergent coronal structures}
\author{P. D. Smith\inst{1} \and D. Tsiklauri\inst{1} \and M. S. Ruderman\inst{2}}
\titlerunning{Enhanced phase mixing of Alfv\'en waves in coronal structures}
\authorrunning{P. D. Smith et al.}
\offprints{\\ Phil Smith, \email{p.d.smith@pgr.salford.ac.uk}}
\institute{Institute for Materials Research, University of Salford, Greater Manchester, M5 4WT, United Kingdom. \and Department of Applied Mathematics, University of Sheffield, Hicks Building, Hounsfield Road, Sheffield, S3 7RH, United Kingdom.}
\date{Received Xerox / Accepted 24/09/2007}

\abstract {} {To explore the solar coronal heating enigma by an analytical and numerical study of the enhanced phase mixing of harmonic Alfv\'en waves propagating in gravitationally stratified coronal structures of varying magnetic field divergence.}{Corrected analytical solutions are derived to model the dissipation of Alfv\'en waves propagating in divergent and stratified coronal structures. These analytical solutions are validated and further explored using a newly developed 2.5D visco-resistive linear MHD code.}{Corrected analytical solutions describing the enhanced phase mixing of Alfv\'en waves in divergent and stratified coronal structures are presented. These show that the enhanced phase mixing mechanism can dissipate Alfv\'en waves at heights less than half that is predicted by the previous analytical solutions. In divergent and stratified coronal structures, the enhanced phase mixing effect occurs only when the ratio of the magnetic and density scale heights, $H_\mathrm{b} / H_\mathrm{\rho} < 2$. The enhanced phase mixing of $0.1$~Hz harmonic Alfv\'en waves propagating in strongly divergent, $H_\mathrm{b} = 5$~Mm, stratified coronal structures, $H_\mathrm{\rho} = 50$~Mm, can fulfill $100\%$ of an active region heating requirement, by generating viscous heating fluxes of $F_\mathrm{H} \approx 2100$~J~m$^{-2}$~s$^{-1}$. The Alfv\'en waves in this configuration are fully dissipated within $20$~Mm, which is six times lower than would occur as a result of standard phase mixing in uniform magnetic fields. This results in the heating length scale, $L_\mathrm{H}$, defined as the height at which $95\%$ of the Alfv\'en wave poynting flux has dissipated, being lowered by a factor of six, to less than half of an active region density scale height. Using the corrected analytical solutions it was found that, for a given wave frequency, the generation of a heating length scale of $L_\mathrm{H} \le 50$~Mm, by enhanced phase mixing in strongly divergent magnetic fields, requires a shear viscosity eight orders of magnitude lower, than required by standard phase mixing in uniform magnetic fields. It was also found that the enhanced phase mixing of observable, $\omega \approx 0.01$~rads~s$^{-1}$, Alfv\'en waves, in strongly divergent magnetic fields, $H_\mathrm{b} = 5$~Mm, can generate heating length scales within a density scale height, $H_\mathrm{\rho} = 50$~Mm, using classical Braginskii viscosity. It is therefore not necessary to invoke anomalous viscosity in corona, if phase mixing takes place in strongly divergent magnetic fields. This study shows that the importance of enhanced phase mixing as a mechanism for dissipating Alfv\'en waves in the solar corona (a stratified and divergent medium), has been seriously underestimated.}{}

\keywords{Magnetohydrodynamics (MHD) - Methods: analytical - Methods: numerical - Sun: corona - Sun: oscillations - Waves}

\maketitle

\section{Introduction}

Phase mixing was proposed by \citet{hp83} as a mechanism for dissipating Alfv\'en waves in the solar corona. They suggested that shear Alfv\'en waves propagating on neighboring magnetic field lines, in inhomogeneous plasmas, would move out of phase with each other. This would then lead to strong gradients, or short length scales, developing perpendicular to the direction of wave propagation. The build up of strong transverse gradients eventually leads to dissipation of the wave's energy via shear viscosity or resistivity, thus heating the plasma.

The ability of phase mixing to make a significant contribution to coronal heating has attracted ample attention of solar physicists. This simple mechanism has been applied to a variety of coronal structures; most notably coronal holes, loops and arcades \citep[see reviews in][]{b91,nu90,nu96,p91}. These studies are supported by observational evidence of Alfv\'en waves, which after being generated by photospheric footprint movements, propagate up through these coronal structures \citep[see][]{i96}. Phase mixing may also contribute to coronal heating through the production of fast and slow magnetosonic waves due to non-linear coupling with Alfv\'en waves \citep[see][]{nrm97,tn02,tna02,tnr03}. Recently \citet{t06a} has shown that this mechanism could lead to significant coronal heating through the generation of parallel electric fields. The main alternative to coronal heating by waves, comes from the well studied reconnection of magnetic fields \citep[see][]{p03,plt03,plh05}.

While Alfv\'en waves contain enough energy to account for the observed one million kelvin temperatures of the corona, their notoriously slow dissipation means that their energy is carried high into the corona. The low classical \citet{b65} values of
shear viscosity and resistivity mean that, even with phase mixing, it is very difficult to dissipate Alfv\'en waves within the few density scale heights, $H_\mathrm{\rho} \approx 50 - 200$~Mm, required for them to be responsible for coronal heating. Numerous studies have therefore sought to lower the height at which Alfv\'en waves dissipate by a variety of mechanisms: localizing the Alfv\'en waves into closed-field regions (e.g. Resonant absorption \citep[see][]{ods94,ods95}), using large (anomalous) values of resistivity and viscosity, and by what we will call the enhanced phase mixing effect. In this latter mechanism, the phase mixing of harmonic Alfv\'en waves propagating in divergent magnetic fields was found to generate dissipation rates far in excess of the standard \citet{hp83} $\exp(-z^3)$ rate (where $z$ is height). Numerous authors have therefore investigated how magnetic field divergence, along with gravitational density stratification, changes the efficiency of phase mixing to dissipate Alfv\'en waves \citep{ptbg97,rnr98,rgr99,dmhia99,dmha00}. A recent study by \citet{awtp07} suggests that $83\%$ of the corona's heating requirement is located in active regions, while \citet{ana00} showed that the heating scale height is required to be less than the active region's density scale height. Therefore, for Alfv\'en waves to be responsible for the majority of coronal heating, a mechanism that can dissipate them in active regions, within a density scale height, is required.

\citet{dmha00} detail the effects of altering the density scale height on the phase mixing of Alfv\'en waves propagating in radially divergent magnetic fields. They concluded that the resultant dissipation can either be enhanced or diminished, depending on the specific configuration of the coronal structure. Analytical solutions for Alfv\'en waves propagating in more general divergent and stratified coronal structures can be found in \citet{rnr98}. In non-stratified divergent coronal structures, they found that the standard dissipation rate of harmonic waves, $\exp(-z^3)$, is replaced by an even faster $\exp(-\exp(z/H_\mathrm{b}))$ rate which depends on the magnetic scale height, $H_\mathrm{b}$. In stratified coronal structures, harmonic Alfv\'en waves were generally found to dissipate slower than the standard rate; although this depended strongly on the specific coronal structure. These two papers used quite different models, making a direct comparison difficult, but they do both generally agree that a diverging magnetic field enhances the phase mixing mechanism, while the gravitational density stratification diminishes it. They also raise a number of important questions: what are the coronal conditions necessary for enhanced phase mixing to occur? How much lower do harmonic Alfv\'en waves dissipate in the corona as a result of enhanced phase mixing? Can the analytical solutions of \citet{rnr98} be confirmed numerically and what is their range of validity? Finally, is enhanced phase mixing a viable mechanism for heating the solar corona to temperatures in excess of one million kelvin?

The aim of this study is to attempt to answer these questions. In the next section we present the model used in our study. In Sect.~3 we correct the analytical solutions of \citet{rnr98}, which describe the phase mixing of Alfv\'en waves in stratified coronal structures with divergent magnetic fields in the WKB approximation. In Sect.~4 we detail the numerical code that was used to solve the linearized MHD equations which describe phase mixing. In Sect.~5 we present our numerical results and compare them to the analytical solutions of Sect.~3. Finally in Sect.~6 we discuss the results and important conclusions that can be drawn from this study.

\section{Analytical Model}

The starting point for our analysis is the system of MHD equations for a cold, incompressible plasma
\begin{equation}
\rho \frac{\partial \vec V}{\partial t} + \rho\left(\vec V \cdot \nabla\right)\vec V = -\frac {1}{\mu_0}\vec B \times \left(\nabla \times \vec B\right) - \nabla\times \left(\rho \nu \nabla \times \vec V\right), \label{momentum}
\end{equation}
\begin{equation}
\frac{\partial \vec B}{\partial t} = \nabla\times\left(\vec V\times\vec B\right)-\nabla\times\left(\eta\nabla\times\vec B\right). \label{induction}
\end{equation}
Here $\vec B$ is the magnetic field, $\vec V$ the plasma velocity, $\rho$ the mass density, $\nu$ the kinematic viscosity, $\eta$ the magnetic diffusivity and $\mu_0$ the magnetic permeability of free space. In the momentum equation, only the shear viscosity term is included, even though the bulk viscosity in the solar corona is many orders of magnitude larger. This approximation is valid as the bulk viscosity is related to compressibility, which is not included in our model and thus does not effect the amplitude of propagating Alfv\'en waves. Indeed, it has been shown through numerical simulations by \citet{ods94} and \citet{eg95}, that, under typical coronal conditions, the bulk viscosity can be neglected in Alfv\'en dissipative layers in comparison with the shear viscosity. The use of the cold plasma approximation in our model applies only to the perturbation quantities of velocity and magnetic field. This approximation holds in the solar corona, as the plasma beta is usually very small ($\beta \approx 0.01$). The equilibrium quantities incorporate thermal physics to account for the vertical density stratification of the corona.

The applicability of the MHD equations for the description of wave propagation in the solar corona, is very often questioned on the grounds that the coronal plasma is collisionless. In this respect, we have to note that it is meaningless to discuss if a plasma is collisional or collisionless before we have specified the characteristic time scale of the problem. A plasma has to be considered as collisionless if the characteristic time scale is much smaller than the ion collision time. Conversely it has to be considered as collisional if the characteristic time scale is much larger than the ion collision time. The collisional time of protons in the solar corona, $\tau_\mathrm{p}$, was calculated by \citet{h85}. He obtained that $\tau_\mathrm{p}$ can vary from about $0.7$~s in active regions to about $7.5$~s near the base of coronal holes. For the particular values of the electron number density ($6.43 \times 10^{14}$~m$^{-3}$) and temperature ($2 \times 10^6$~K) used in our numerical calculations we obtain $\tau_\mathrm{p} \approx 3$~s. In the following calculations, we consider waves with periods larger than or equal to $10$~s, so that the use of the MHD description is fully justified. It is also known that the relatively strong magnetic fields present in the corona, inhibit cross-field particle motions, while wave particle interactions are known to impede particle motions even along the field. These have the combined effect of localizing the particle interactions in all directions, thereby enabling a fluid description to apply \citep[see][]{pf00}. Given this, there have been various attempts to include effects beyond MHD into coronal studies; \citet{tss05b} included ion and electron kinetic effects, \citet{ov07} considered multiple species of ions, while \citet{t07} has included the effects of two fluid species.

In what follows we use the Cartesian coordinates $x,y,z$ to represent respectively the transverse coordinate, the ignorable coordinate ($\partial / \partial y = 0$) and the height in the corona. We neglect the magnetic field diffusion ($\eta = 0$). We assume a static ($\vec V_0 = 0$) two dimensional ($B_\mathrm{0y} = 0$) equilibrium, where the subscript '0' indicates an equilibrium quantity. We then linearize Eqs.~(\ref{momentum})--(\ref{induction}) and consider perturbations in the form of linearly polarized Alfv\'en waves, so that only the y-components of the perturbations of $\vec V$ and $\vec B$ are non-zero. This results in the following system of linear equations:
\begin{equation}
\rho_0 \frac{\partial V_\mathrm{y}}{\partial t} = -\frac {1}{\mu_0} \left( \vec B_0 \cdot \nabla \right) B_\mathrm{y} + \nabla \cdot (\rho_0 \nu \nabla V_\mathrm{y}), \label{mhd1}
\end{equation}
\begin{equation}
\frac {\partial B_\mathrm{y}}{\partial t} = \left( \vec B_0 \cdot \nabla \right) V_\mathrm{y}. \label{mhd2}
\end{equation}
Eliminating $B_\mathrm{y}$ from Eqs.~(\ref{mhd1})--(\ref{mhd2}) gives the diffusive wave equation
\begin{equation}
\rho_0 \frac{\partial^2V_\mathrm{y}}{\partial t^2} = -\frac {1}{\mu_0} \left( \vec B_0 \cdot \nabla \right)^2 V_\mathrm{y} + \nabla \cdot \left(\rho_0 \nu \nabla \frac{\partial V_\mathrm{y}}{\partial t}\right). \label{wave_eqn}
\end{equation}

Since the equilibrium magnetic field is two-dimensional and divergence-free, it can be expressed in terms of the flux function $\psi$,
\begin{eqnarray}
B_{\mathrm{0x}} = -B_{00} \frac{\partial \psi}{\partial z} , && B_{\mathrm{0z}} = B_{00} \frac{\partial \psi}{\partial x},
\label{b0xz}
\end{eqnarray}
where $B_{00}$ is the magnitude of the magnetic field at the coordinate origin ($x = z = 0$). An individual magnetic field line is described by $\psi = \mathrm{const}$. We can introduce a second function, $\phi$, satisfying $\nabla \psi \cdot \nabla \phi = 0$, and consider $\psi$ and $\phi$ as new curvilinear coordinates in the $xz$-plane. Thus Eq.~(\ref{wave_eqn}) can be re-written as
\begin{eqnarray}
\sigma \frac{\partial^2 V_\mathrm{y}}{\partial t^2} & = & V_\mathrm{A0}^2 J \frac{\partial}{\partial \phi} J \frac{\partial V_\mathrm{y}}{\partial \phi} \nonumber \\
&& + J \frac{\partial}{\partial t} \left[ \frac{\partial}{\partial \psi} \left( \sigma \nu J h_\phi^2 \frac{\partial V_\mathrm{y}}{\partial \psi} \right) + \frac{\partial}{\partial \phi} \left( \sigma \nu J h_\psi^2 \frac{\partial V_\mathrm{y}}{\partial \phi} \right) \right], \label{wave_eqn2}
\end{eqnarray}
where $\sigma = \rho_0 / \rho_{00}$ is the dimensionless density, $\rho_{00}$ is the density at the coordinate origin, $V_\mathrm{A0}$ is the Alfv\'en speed at the coordinate origin given by $V_\mathrm{A0} = B_{00}/(\mu_0\rho_{00})^{1/2}$, $J$ is the Jacobian of the coordinate transformation,
\begin{equation}
J = \frac{\partial \psi}{\partial x}\frac{\partial \phi}{\partial z} - \frac{\partial \psi}{\partial z}\frac{\partial \phi}{\partial x},
\label{jacobian}
\end{equation}
and $h_\psi$ and $h_\phi$ are the scale factors given by
\begin{eqnarray}
h_\psi  = \left[ \left( \frac{\partial z}{\partial \psi} \right)^2  + \left( \frac{\partial x}{\partial \psi} \right)^2 \right]^{1/2}, && h_\phi  = \left[ \left( \frac{\partial z}{\partial \phi} \right)^2  + \left( \frac{\partial x}{\partial \phi} \right)^2 \right]^{1/2}.
\label{scale_factors}
\end{eqnarray}
Note that our definition of these scale factors differs from those of \citet{rnr98}, resulting in a slightly different form of Eq.~(\ref{wave_eqn2}). The scale factors used by \citet{rnr98}, $h_{\psi\mathrm{r}}$ and $h_{\phi\mathrm{r}}$, are related to ours by $h_{\psi\mathrm{r}} = J h_\phi^2$ and $h_{\phi\mathrm{r}} = J h_\psi^2$. In either case, both definitions of the scale factors lead to identical general solutions.

\section{Analytical Solutions}

\citet{rnr98} obtained an analytical solution of Eq.~(\ref{wave_eqn2}) in the WKB approximation. Their approach is valid when the following assumptions are satisfied
\begin{enumerate}
\item The ratio of the characteristic scales in $x$ and $z$-directions is small, $x_0/H \ll 1$;
\item The wavelength is $\leq x_0$;
\item The characteristic scale of damping is $\leq H$.
\end{enumerate}
Here $H$ is the smaller of the two quantities, the characteristic scale of the magnetic field variation, $H_\mathrm{b}$, and the density scale height $H_\rho$.

We verified the calculations of \citet{rnr98}, and found that they are valid up to their Eq.~(36). However there is an error in their Eq.~(37). The correct equation should be
\begin{equation}
I(x,z) = \int_{0}^{z}\frac {\partial x}{\partial \psi}\frac {1}{V_\mathrm{A}^2} \frac{\partial V_\mathrm{A}}{\partial x} dz', \label{pdsgen2}
\label{i_missing}
\end{equation}
where the Alfv\'en speed is given by
\begin{equation}
V_\mathrm{A} = \frac{B_0}{\left( \mu_0 \rho_0 \right)^{1/2}}.
\label{va}
\end{equation}
\citet{rnr98} missed the factor $\partial x / \partial \psi$ in their expression for $I(x,z)$. If we take $V_{\mathrm{y}} \propto \exp (-i \omega t)$, then the evolution of the wave velocity is given by
\begin{equation}
V_\mathrm{y}(x,z) = \sigma^{-1/4} W(x)e^{-\Lambda(x,z)},
\label{amplitude}
\end{equation}
where
\begin{equation}
\Lambda(x,z) \approx \frac{\omega^2}{2B_{00}^2} \int_0^z \frac{ \nu(x,z') \vec B_0^2(x,z') I^2(x,z')}{V_\mathrm{A}(x,z')} dz',
\label{pdsgen1}
\end{equation}
and $W(x)$ is determined by the boundary condition at $z = 0$. For a detailed derivation of this general solution, see Appendix A.

Now, using the corrected expression for $I(x,z')$ we reconsider three particular examples of equilibria studied by \citet{rnr98}. Following them we assume that $\nu = $ const. The first equilibrium they studied was a uniform vertical magnetic field in an isothermal atmosphere, so that $B_\mathrm{0x} = 0$, $B_\mathrm{0z} = B_{00}$ and $\rho_0(x,z) = \hat{\rho}_0(x) e^{-z/H_\mathrm{\rho}}$. Where again, the term isothermal applies to the equilibrium quantities only. For that specific configuration $\psi = x$, $\phi = z$, $\partial x / \partial \psi = 1$ and therefore the expression for $I(x,z')$ given by \citet{rnr98} coincides with Eq.~(\ref{pdsgen2}). Hence, in this particular case, the analytical solution obtained by Ruderman et al. is correct.

The two other cases studied by \citet{rnr98} were equilibria with (i) constant density along $z$ (non-stratified) and an exponentially diverging magnetic field, and (ii) constant Alfv\'en speed along $z$. We consider a more general equilibrium consisting of an exponentially diverging magnetic field in an isothermal stratified atmosphere. In this case the magnetic flux function is given by
\begin{equation}
\psi(x,z) = H_\mathrm{b} e^{-z/H_\mathrm{b}}\sin(x/H_\mathrm{b}) \label{magx},
\end{equation}
so that the magnetic field is determined by
\begin{eqnarray}
B_\mathrm{0x}= B_{00} e^{-z/H_\mathrm{b}} \sin \left( x/H_\mathrm{b} \right), && B_\mathrm{0z}= B_{00} e^{-z/H_\mathrm{b}} \cos \left( x/H_\mathrm{b} \right).
\label{flux_functions}
\end{eqnarray}
The orthogonal curvilinear coordinate $\phi$ is given by
\begin{equation}
\phi(x,z) = H_\mathrm{b} e^{-z/H_\mathrm{b}} \cos (x/H_\mathrm{b}).
\end{equation}
The expression for the equilibrium density is
\begin{equation}
\rho_0(x,z) = \hat{\rho}_0(x) e^{-z/H_\mathrm{\rho}}.
\label{densityz}
\end{equation}
The exact form of the missing factor from Eq.~(\ref{i_missing}) is given by $\partial x / \partial \psi = \sec \left( x/H_\mathrm{b} \right) \exp(z/H_\mathrm{b})$. In the following solutions, we use the simplified form $\partial x / \partial \psi \approx \exp(z/H_\mathrm{b})$, since $\sec \left( x/H_\mathrm{b} \right) \approx 1$.
Then, using Eq.~(\ref{pdsgen2}), we obtain
\begin{equation}
I(x,z) = \frac{H_\rho H_\mathrm{b} \left[ 1 - \exp ( 2z/H_\mathrm{b}  - z/2H_\rho ) \right]}{V_{A0} \hat{\rho}_0^{1/2} \rho_{00}^{1/2} \left( 4H_\rho - H_\mathrm{b} \right)}  \frac{\partial \hat{\rho}_0}{\partial x}.
\label{cs4I}
\end{equation}
Substituting Eq.~(\ref{cs4I}) into Eq.~(\ref{pdsgen1}) we arrive at
\begin{eqnarray}
\Lambda (x,z) & \approx & \bar \Lambda (x) \left[ \frac{1 - \exp( 3z/H_\mathrm{b} - 3z/2H_\rho)}{3\left( H_\mathrm{b} - 2H_\rho \right)} \right. \nonumber \\
& & \quad \quad + \quad \frac{1 - \exp( z/H_\mathrm{b} - z/H_\rho)}{H_\rho - H_\mathrm{b}} \nonumber \\
& &  \left. \quad \quad + \quad \frac{1 - \exp( - z/2H_\rho - z/H_\mathrm{b})}{H_\mathrm{b} + 2H_\rho} \right],
\label{pds4_l}
\end{eqnarray}
\begin{equation}
\bar \Lambda (x) = \frac{\nu \omega^2}{V_\mathrm{A0}^3 \hat \rho_0^{1/2} \rho_{00}^{3/2}} \frac{H_\rho^3 H_\mathrm{b}^3}{\left( 4H_\rho - H_\mathrm{b} \right)^2} \left( \frac{\partial \hat \rho_0}{\partial x}\right)^2.
\label{pds4_lb}
\end{equation}
Clearly, Eqs.~(\ref{pds4_l})--(\ref{pds4_lb}) are only valid when $H_\mathrm{b} \ne H_\rho$, $H_\mathrm{b} \ne 2H_\rho$ and $H_\mathrm{b} \ne 4H_\rho$. In what follows we assume that these inequalities hold.

By considering a non-stratified coronal structure permeated by a uniform magnetic field, where $H_\rho \to \infty$ and $H_\mathrm{b} \to \infty$, we can obtain the \citet{hp83} standard phase mixing solution; Eqs.~(\ref{pds4_l})--(\ref{pds4_lb}) reduce to
\begin{equation}
\Lambda(x,z) = \frac{ \nu \omega^2}{6V_\mathrm{A}^5}  \left(\frac{\partial V_\mathrm{A}}{\partial x} \right)^2 z^3 .
\label{hp_l}
\end{equation}

To obtain the equilibrium which represents a non-stratified divergent coronal structure, we take $H_\rho \to \infty$. Then Eqs.~(\ref{pds4_l})--(\ref{pds4_lb}) reduce to
\begin{equation}
\Lambda (x,z) \approx \bar \Lambda (x) \left( 1 + 3e^{- z/H_\mathrm{b}} \right) \left( e^{z/H_\mathrm{b}} - 1 \right)^3,
\label{pds2_l}
\end{equation}
\begin{equation}
\bar \Lambda(x)  = \frac{ \nu \omega^2 H_\mathrm{b}^3}{96 V_\mathrm{A0}^3 \hat \rho_0^{1/2} \rho_{00}^{3/2} } \left( \frac{\partial \hat \rho_0}{\partial x} \right)^2.
\label{pds2_lb}
\end{equation}
We see that, when $z$ is a few times larger than $H_\mathrm{b}$, the wave amplitude is proportional to $\exp\left(-\bar \Lambda e^{3z/H_\mathrm{b}}\right)$, so that wave damping occurs much faster than in the one-dimensional case studied by \citet{hp83}, with the characteristic damping length proportional to $\ln \nu$. Qualitatively this result is in agreement with the \citet{rnr98} solution, however the exact expressions for $\Lambda$ and $\bar \Lambda$ that we obtained are different from their corresponding expression (see their Eqs.~(45)--(46)).

To obtain the equilibrium with constant Alfv\'en speed we take $H_\rho \to \frac{1}{2} H_\mathrm{b}$. Then Eqs.~(\ref{pds4_l})--(\ref{pds4_lb}) reduce to
\begin{equation}
\Lambda (x,z) \approx \bar \Lambda (x)\left[ \left( e^{-z/H_\mathrm{b}}  - 3 \right)\left( 1 - e^{- z/H_\mathrm{b}} \right) + \frac{2z}{H_\mathrm{b}} \right],
\label{pds3_l}
\end{equation}
\begin{equation}
\bar \Lambda (x) = \frac{\omega^2 \nu H_\mathrm{b}^3}{16V_\mathrm{A} \hat \rho_0^2} \left( \frac{\partial \hat \rho_0}{\partial x} \right)^2.
\label{pds3_lb}
\end{equation}
where $V_\mathrm{A}(x)$ is no longer a function of $z$, as the divergence of the magnetic field exactly balances the effect of stratification. We see that, when $z \gtrsim H_\mathrm{b}$, $\Lambda \propto z$ so that wave damping occurs much slower than in the \citet{hp83} equilibrium where $\Lambda \propto z^3$. Using the incorrect general solution for $\Lambda$, \citet{rnr98} obtained that, in the equilibrium with constant Alfv\'en speed, the wave amplitude tends to a non-zero constant as $z \to \infty$, whereas our corrected analytical solution shows that this is clearly not the case. See Appendix B for detailed derivations of these equilibrium solutions.

In the following sections we will compare our numerical calculations to our corrected general solution as well as to the original \citet{rnr98} general solution. This is done solely to emphasize the importance of the correction presented in this study. The authors note that, in spite of an error in the calculations of \citet{rnr98}, their results remain qualitatively correct. In particular, their conclusion that magnetic field divergence can strongly enhance phase mixing, is in agreement with the conclusions of this study.

\section{The Numerical Model}

\begin{figure}[t]
\resizebox{\hsize}{!}{\includegraphics{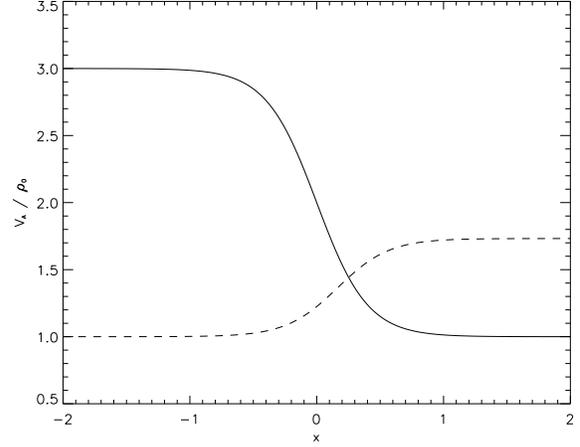}}
\caption{Normalized density (solid) and Alfv\'en velocity (dashed)
as a function of the transverse coordinate $x$, which in dimensional units is given in Mm.}
\label{density&va}
\end{figure}

We developed and tested a parallelized 2.5D visco-resistive linear MHD code, consisting of a centered finite difference scheme (6th order in space), combined with a Runge-Kutta time step (4th order in time). The 2.5D nature of the code means that all variables are function of $x$ and $z$ only, however $\vec B$ and $\vec V$ have all three components. We then used this code to solve Eqs.~(\ref{mhd1})--(\ref{mhd2}), written in dimensionless form. We used reference length, $l$, time, $\tau$, velocity, $u$, number density, $N$ and mass density, $R$, to introduce the dimensionless quantities. For a typical coronal hole $l = 1$~Mm, $\tau = 1$~s, $u = 1$~km~s$^{-1}$, $N = 1$~m$^{-3}$ and $R = 1$~kg~m$^{-3}$. From here on, unless specified, all quantities are given in dimensionless form.

In our numerical calculations we modeled various coronal structures, determined by Eqs.~(\ref{flux_functions})--(\ref{densityz}), along with
\begin{equation}
\hat{\rho}_0(x) = \bar{\rho}_0\left[1+\tanh\left(\frac{x+5}{0.4}\right)-\tanh\left(\frac{x}{0.4}\right)\right].
\label{densityx}
\end{equation}
where $\bar{\rho}_0$ is the background density at $x = 2$, $z = 0$, which, using the mean ion mass (measured in proton mass units) $\mu = 1.27$ and the electron number density $n_0 = 6.43 \times 10^{14}$, gives $\bar{\rho}_0 \approx 1.4 \times 10^{-12}$. The variations of $\hat{\rho}_0(x)$ and the Alfv\'en speed across $x$ at $z = 0$, are shown in Fig.~\ref{density&va}. This figure shows a three fold increase in density, along with a corresponding $1/\sqrt{3}$ decrease in Alfv\'en velocity from $\approx 750$ to $\approx 350$, as we move across the density boundary. This is in agreement with observations of solar coronal plume boundaries \citep[see][]{dhgtphhh97}. Fig.~\ref{density&va} also shows the density boundary has a characteristic scale of variation, or half width, of $x_0 \approx 1$. For large magnetic scale heights, $H_\mathrm{b} \ge 700$, Eq.~(\ref{densityx}) represents the boundary of a coronal plume. For lower magnetic scale heights, where $H_\mathrm{b} \le 100$, we use it to represent the boundary of a divergent coronal loop.

The density scale height was fixed at $H_\mathrm{\rho} = 50$ in all numerical calculations with stratified equilibria. In dimensional form $H_\mathrm{\rho} = 50$~Mm, which corresponds to a temperature equal to $2$~MK \citep[see][]{a04}. The transverse density function, Eq.~(\ref{densityx}), was chosen such that the centers of the density boundary and magnetic field divergence coincide at $x = 0$. In Fig.~2 the curvilinear coordinate system is shown, where the white box shows the size of the computation domain.

\begin{figure}[t]
\resizebox{\hsize}{!}{\includegraphics{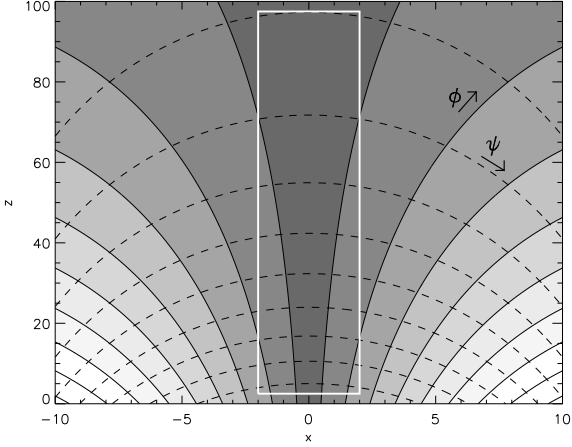}} \caption{Magnetic surfaces defined by curvilinear coordinates $\psi(x,z)$ (dashed) and
$\phi(x,z)$ (solid). The width of the computational domain is shown by the white box.} \label{psiphi}
\end{figure}

The plasma, which initially is at rest, is driven at $z = 0$, by
\begin{eqnarray}
B_\mathrm{y} =  A_0 B_{00} \sin\left(\omega t\right) \left\{ 1-\exp\left[-\left(t/t_0\right)^3\right] \right\},
\label{by_driver}
\end{eqnarray}
\begin{eqnarray}
V_\mathrm{y} = -A_0 \bar{V}_{\mathrm{A}} \sin\left(\omega t\right) \left\{1-\exp \left[-\left(t/t_0\right)^3\right] \right\},
\label{vy_driver}
\end{eqnarray}
where $B_{00}$ is the magnitude of equilibrium magnetic field at $x = z = 0$, $\bar{V}_{\mathrm{A}}$ is the Alfv\'en velocity at the driving boundary, $A_0 = 0.1$ is the dimensionless driving amplitude, $t$ is the simulation time and $t_0 = 100$. \citet{hbw02} describe how the leading and trailing edges of a finite wave train develop into gaussian pulses that dissipate algebraically according to $z^{-3/2}$. Since these leading edge pulses dissipate at a much slower rate than the exponential rate we were seeking, it was important to minimize their effect on our numerical calculations. This was achieved by ramping up $B_\mathrm{y}$ and $V_\mathrm{y}$ over $20$ wavefronts, by including the exponential dampening terms $\exp\left[-\left(t/t_0\right)^3\right]$ seen in the boundary conditions of Eqs.~(\ref{by_driver})--(\ref{vy_driver}). This minimizes the development of leading edge gaussian pulses and enabled the numerical calculations to reach a steady state before tracking of the wave amplitude begins.

In all calculations the kinematic viscosity was $\nu = 5 \times10^{-5}$ (anomalous), while the wave frequency was $f = 0.1$ ($\omega = 0.2\pi$). In dimensional form these correspond to  $\nu = 5 \times10^7$~m$^2$~s$^{-1}$ and $f = 0.1$~Hz. See Sect.~6 for a discussion on the plausibility of these parameters.

The size of the numerical domain was $-2~\le~x~\le~2$, $0~\le~z~\le~100$. We used a combination of zero-gradient ($x = \pm2$), line-tied ($z = 0$) and masked ($z = 100$) boundary conditions. Masking allows waves to propagate through the upper boundary ($z = 100$) without reflection. To use this method we extended the computational domain in the $z$-direction by 50\% to create a masking region where all perturbations are exponentially damped. The corresponding numerical resolution, which includes the masking region, was $400~\times~7500$. The numerical convergence test was carried out by repeating numerical calculations for the coronal structure with the most divergent magnetic field ($H_\mathrm{b} = 5$, $H_\mathrm{\rho} = 50$), with double numerical resolution; $800~\times~15\,000$. Each numerical calculation was run for a time $t = 1000$, on a dual-core Intel Xeon processor for approximately two weeks (depending on $H_\mathrm{b}$), while the convergence check was run on $8$ AMD Opteron processors for a period of one month.

\section{Numerical Results}

In this section we compare the analytical solutions of Sect.~3 to numerical calculations of Alfv\'en waves propagating in divergent stratified coronal structures. We divided our numerical calculations into two regimes: weakly divergent ($H_\mathrm{b} = 40,100,700,\infty$) and strongly ($H_\mathrm{b} = 5,10$) divergent coronal structures. In both regimes, coronal structures both with and without density stratification ($H_\rho = 50$) were considered.

\subsection{Variation of wavelength}

\begin{figure}[t]
\resizebox{\hsize}{!}{\includegraphics{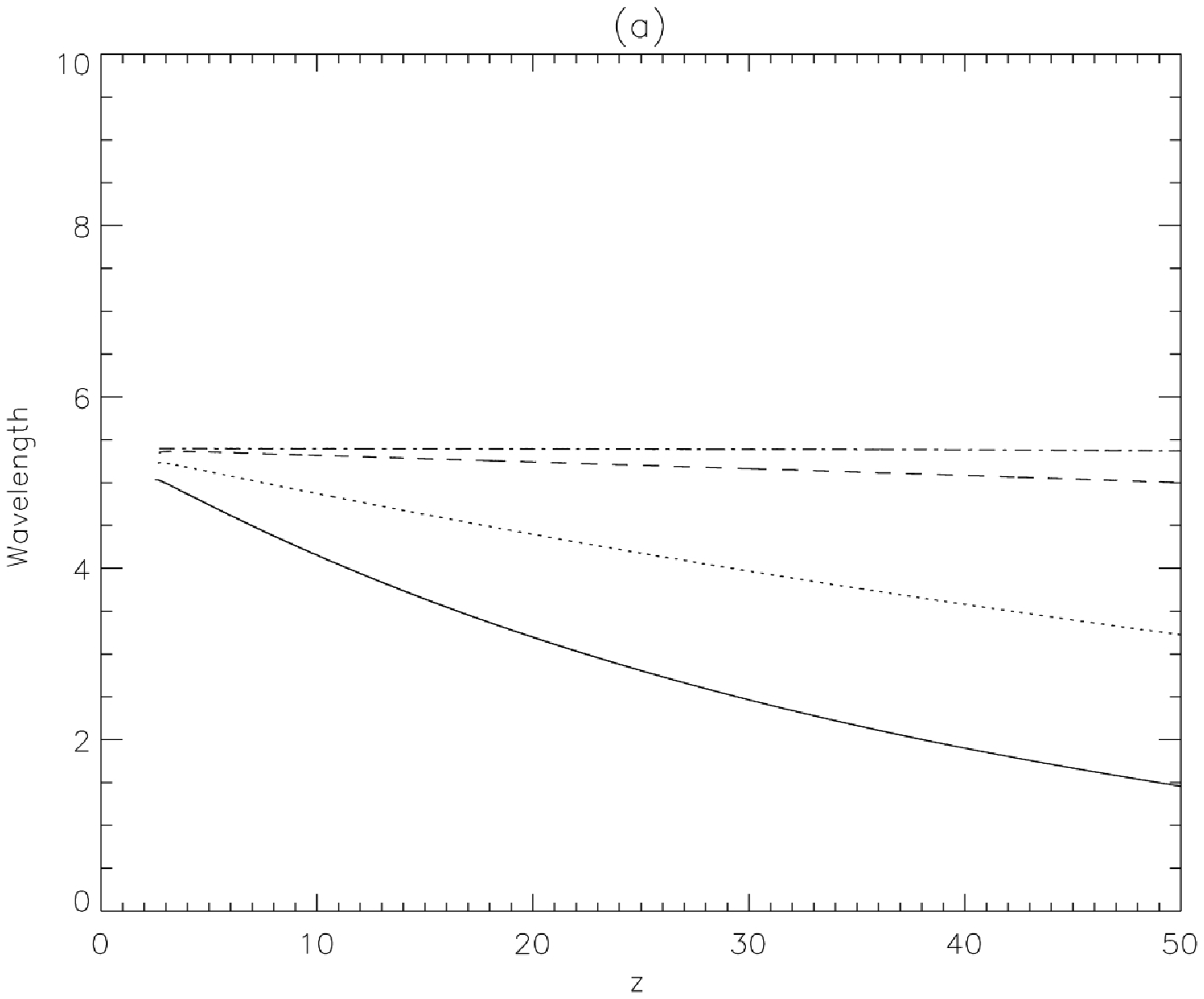}} \resizebox{\hsize}{!}{\includegraphics{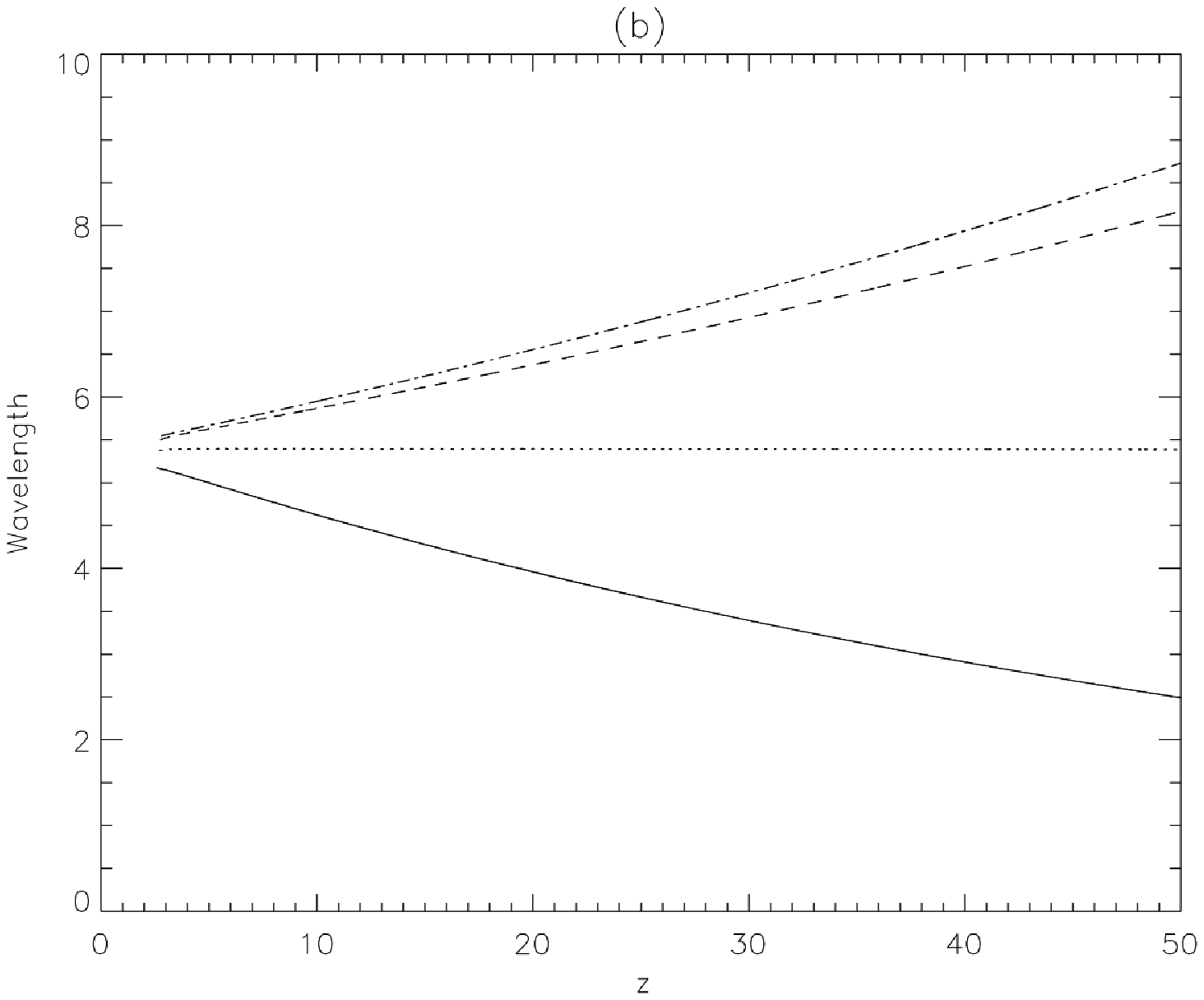}} \caption{Variation of
wavelength with height for (a) non-stratified, $H_\mathrm{\rho} = \infty$, and (b) stratified, $H_\mathrm{\rho} = 50$, weakly divergent coronal structures along $x = 0$. Solid, dotted, dashed and dashed-dotted lines correspond to $H_\mathrm{b} = 40$, $H_\mathrm{b} = 100$, $H_\mathrm{b} = 700$, and $H_\mathrm{b} = \infty$ respectively. In dimensional units, the wavelength, $\lambda_{||}$, and height, $z$, are given in Mm.}
\label{wavelengths}
\end{figure}

\begin{figure}[t]
\resizebox{\hsize}{!}{\includegraphics{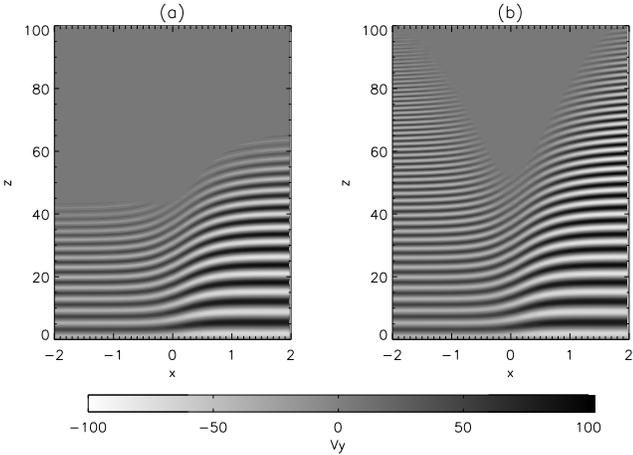}} \caption{An example enhanced phase mixing simulation at times (a) $t = 200$ and (b) $t = 600$, for a weakly divergent, $H_\mathrm{b} = 40$, stratified, $H_\mathrm{\rho} = 50$, coronal structure. In dimensional units, $x$ and $z$ are given in Mm, while $V_\mathrm{y}$ is given in km~s$^{-1}$.} \label{contour}
\end{figure}

\begin{figure*}[t]
\includegraphics[width=0.5\linewidth]{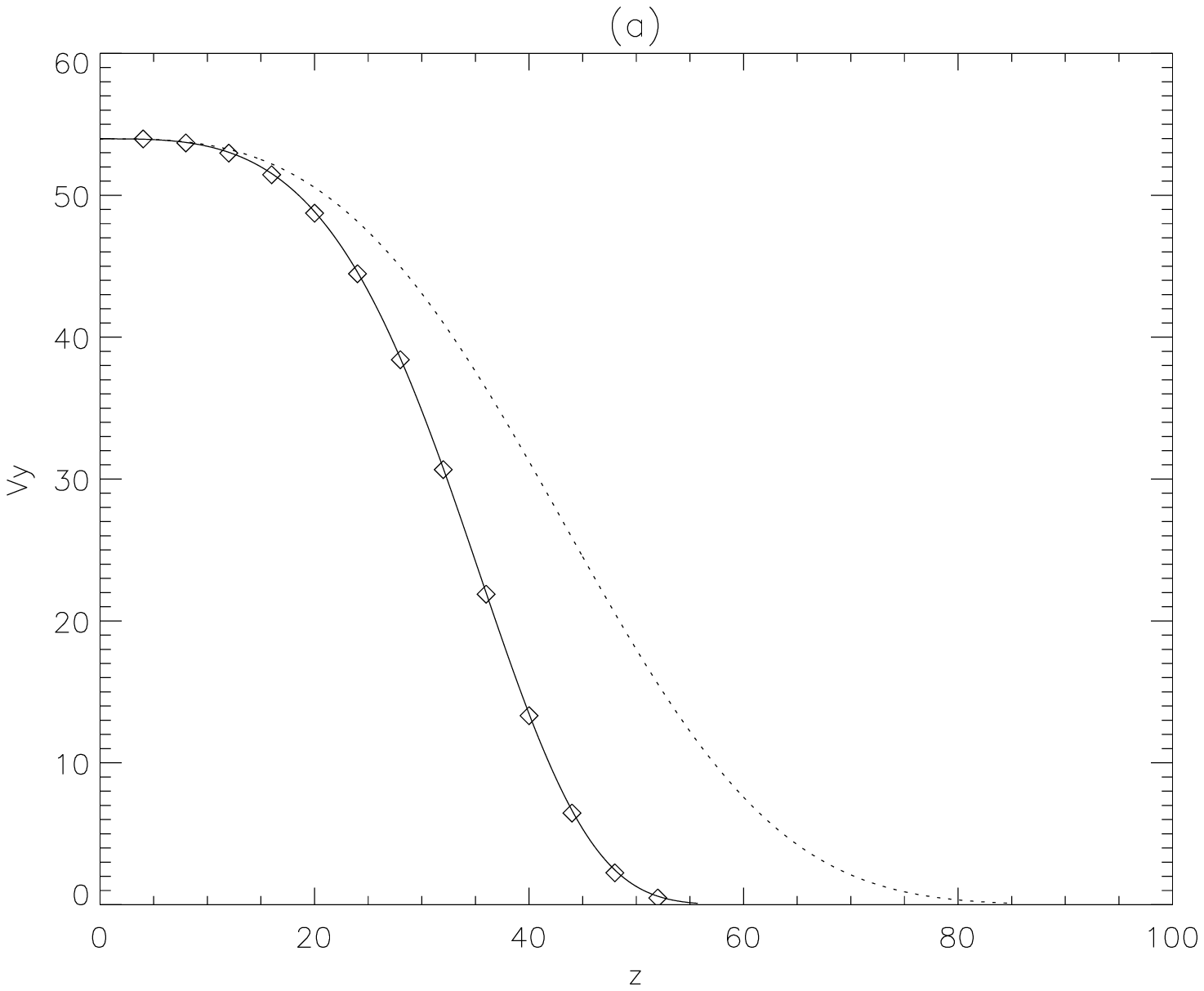}
\includegraphics[width=0.5\linewidth]{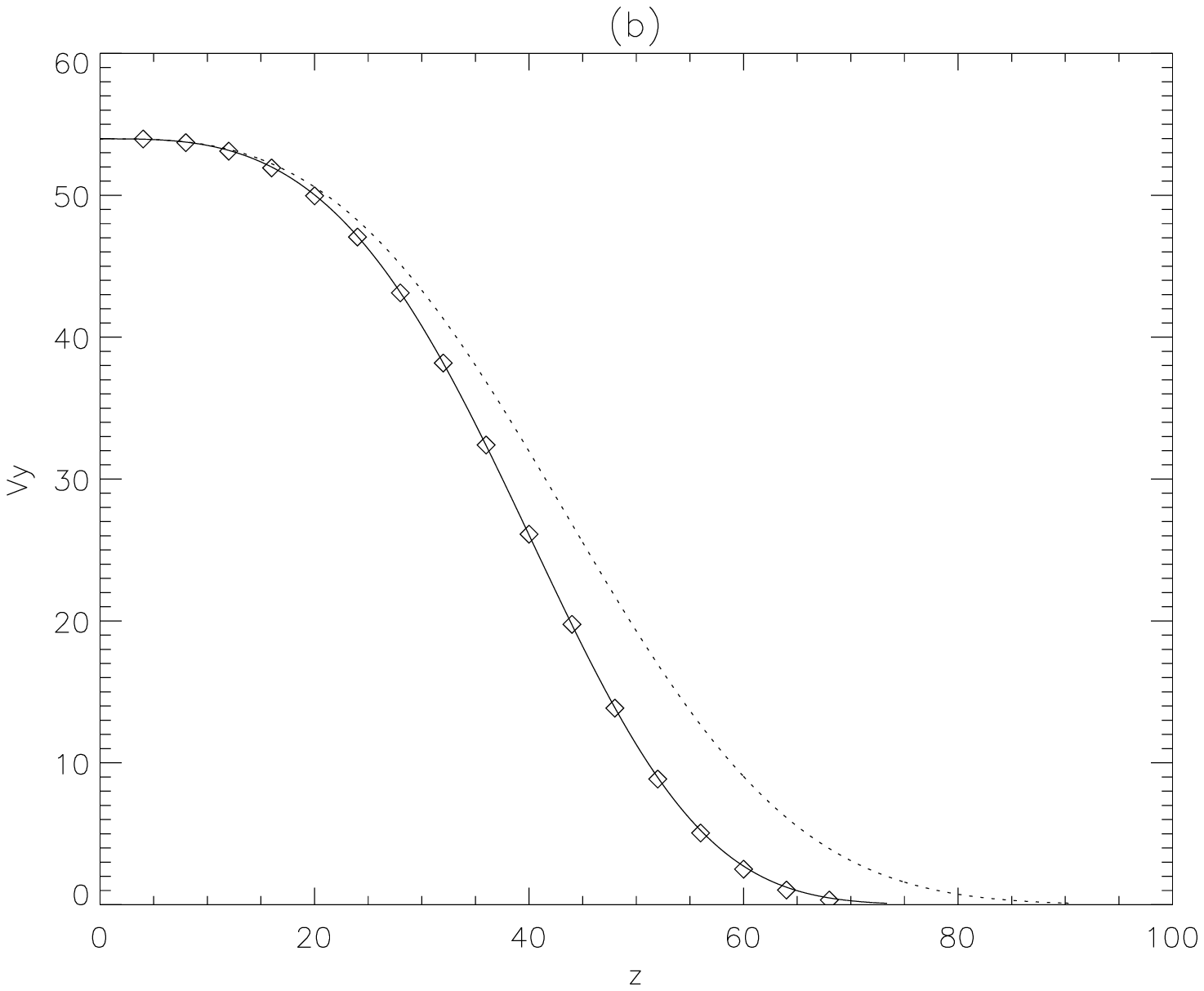}
\includegraphics[width=0.5\linewidth]{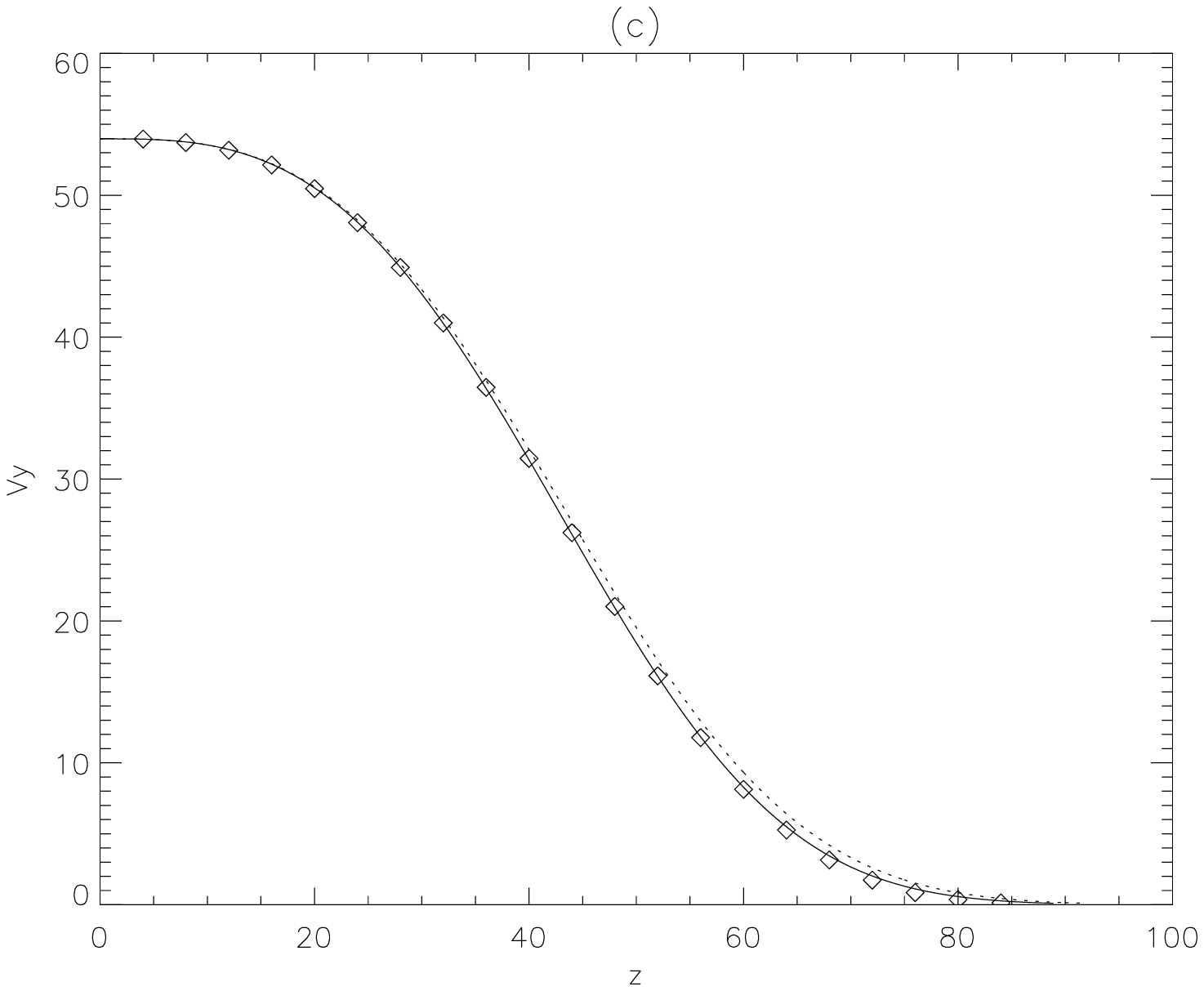}
\includegraphics[width=0.5\linewidth]{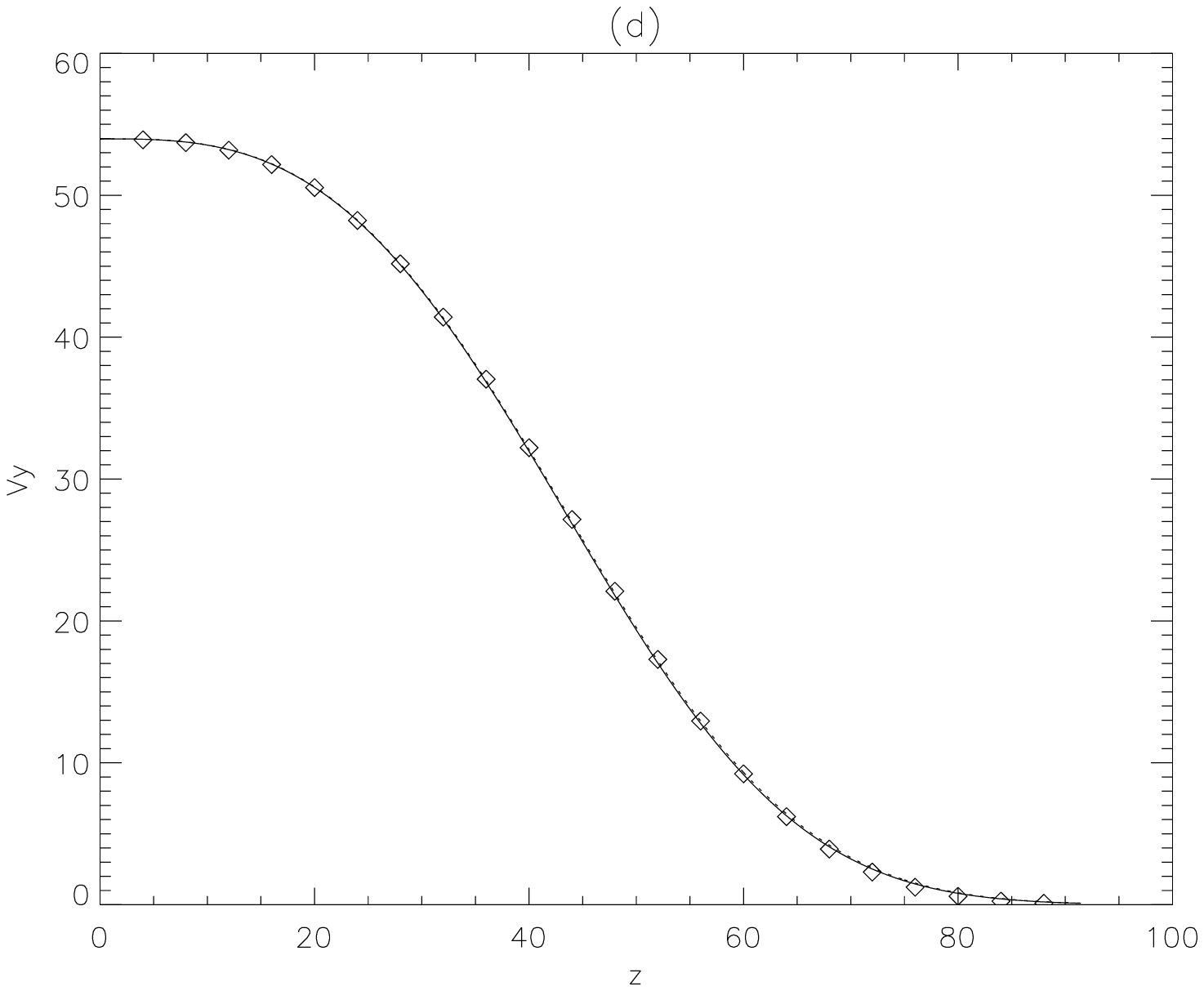}
\caption{Alfv\'en wave amplitude as a function of height along $x = 0$ for non-stratified, $H_\mathrm{\rho} = \infty$, coronal structures with magnetic field divergence (a) $H_\mathrm{b} = 40$, (b) $H_\mathrm{b} = 100$, (c) $H_\mathrm{b} = 700$, (d) $H_\mathrm{b} = \infty$ along $x = 0$. The solid and dotted lines shows calculations from the corrected and original \citet{rnr98} solution respectively, while the diamonds shows the numerical calculations. The dimensional units of amplitude, $V_\mathrm{y}$, and height, $z$, are respectively km~s$^{-1}$ and Mm.}\label{amp_hpinf}
\end{figure*}
\begin{figure*}[t]
\includegraphics[width=0.5\linewidth]{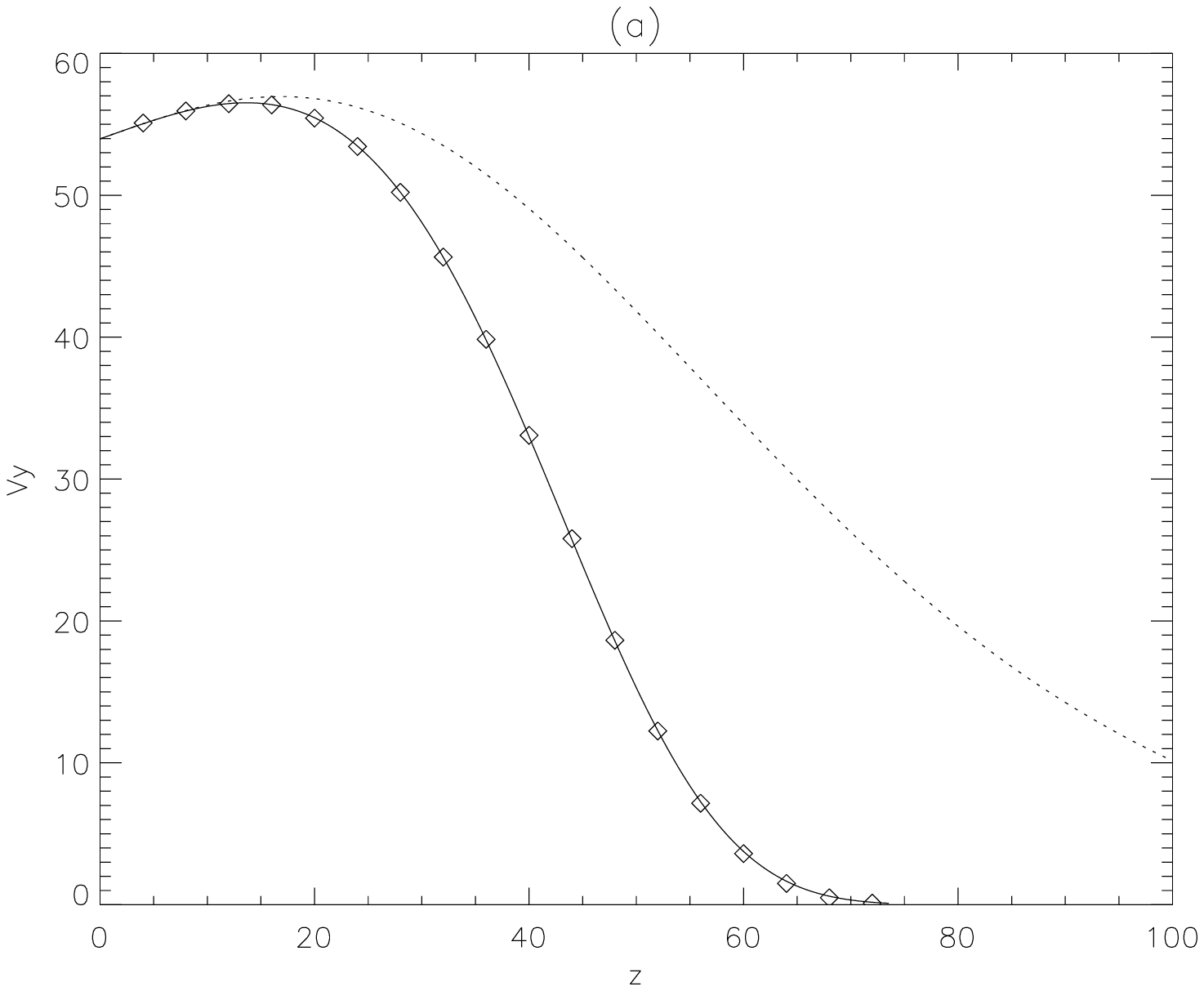}
\includegraphics[width=0.5\linewidth]{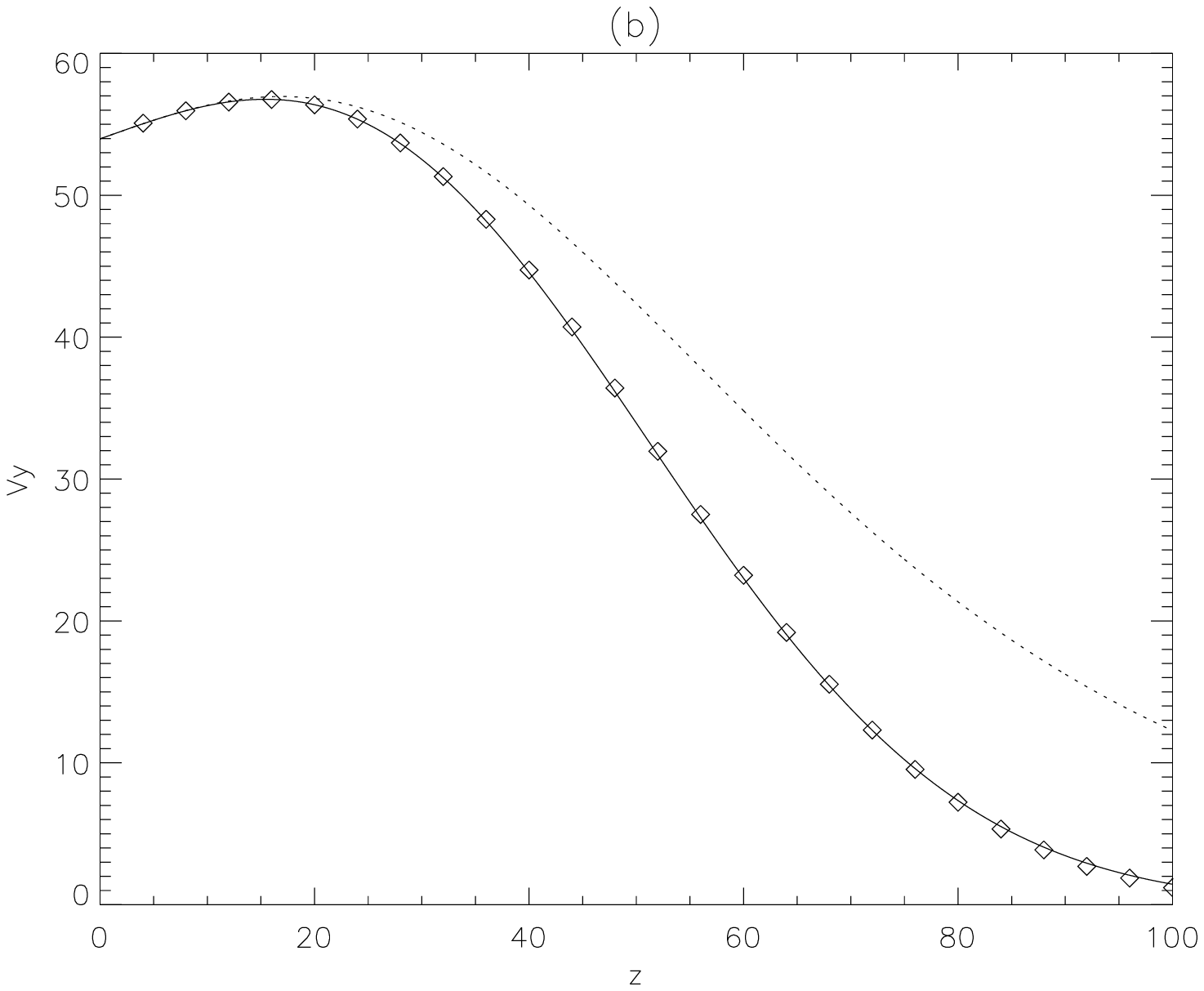}
\includegraphics[width=0.5\linewidth]{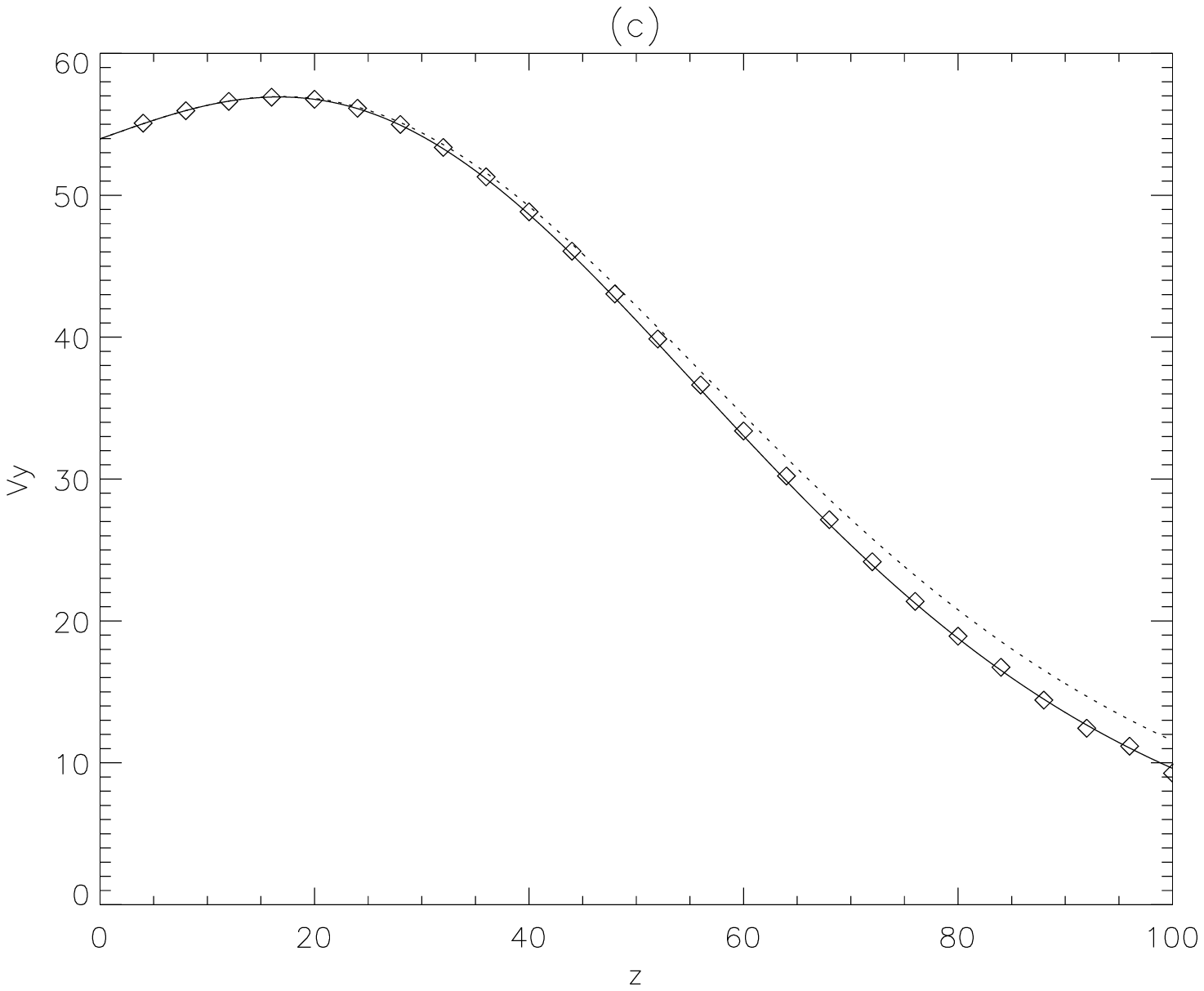}
\includegraphics[width=0.5\linewidth]{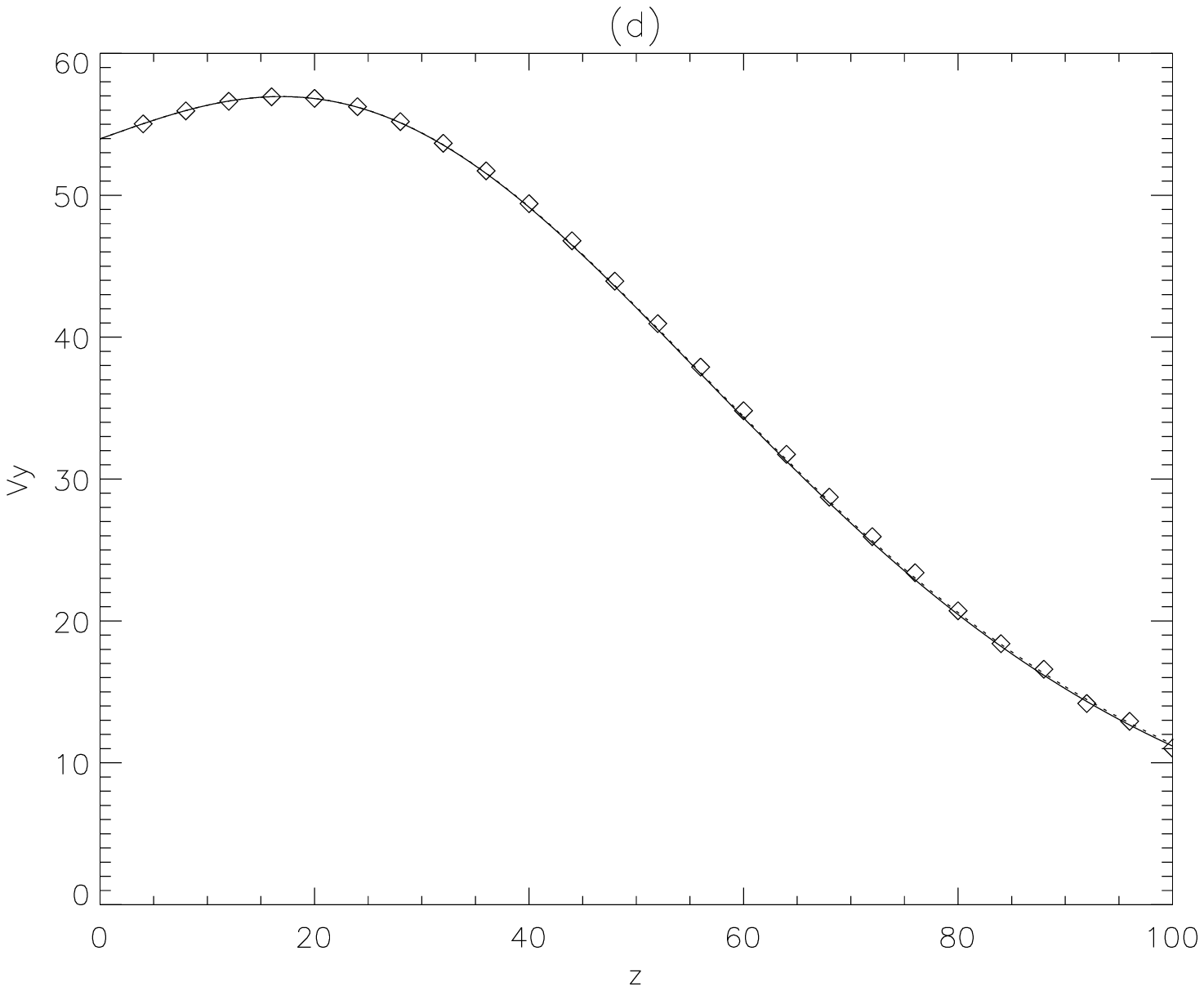}
\caption{Alfv\'en wave amplitude as a function of height along $x = 0$ for stratified, $H_\mathrm{\rho} = 50$, coronal structures of magnetic field divergence (a) $H_\mathrm{b} = 40$, (b) $H_\mathrm{b} = 100$, (c) $H_\mathrm{b} = 700$, (d) $H_\mathrm{b} = \infty$ along $x = 0$. The solid and dotted lines shows calculations from the corrected and original \citet{rnr98} solution respectively, while the diamonds shows the numerical calculations. The dimensional units of amplitude, $V_\mathrm{y}$, and height, $z$, are respectively km~s$^{-1}$ and Mm.}\label{amp_hp50}
\end{figure*}

We begin by examining Fig.~\ref{wavelengths}(a) and Fig.~\ref{wavelengths}(b), which show the variation of wavelength, $\lambda_{||}$, with $z$ for a range of magnetic scale heights, in both stratified and non-stratified weakly divergent coronal structures. The wavelength varies with height as the Alfv\'en velocity, Eq.~(\ref{va}), depends on the local magnetic field strength and local plasma density, which were both chosen to decrease exponentially with height according to their respective scale heights, $H_\mathrm{b}$ and $H_\mathrm{\rho}$. In the standard \citet{hp83} phase mixing solution, Eq.~(\ref{hp_l}), the Alfv\'en velocity, and hence wavelength, is a function of $x$ only; it does not change with height. Whereas in our phase mixing solution given by Eqs.~(\ref{pds4_l})--(\ref{pds4_lb}), the Alfv\'en velocity and wavelength are functions of both $x$ and $z$. A divergent magnetic field will therefore cause the Alfv\'en velocity, and thus the wavelength, to decrease with height. As a result stronger transverse gradients form, leading to greater viscous dissipation than would otherwise occur by standard phase mixing alone. We therefore define enhanced phase mixing to mean phase mixing occurring in divergent magnetic fields which results in stronger dissipation, due to a reduction in the wavelength. In Fig.~\ref{contour}, we show an example simulation of the enhanced phase mixing effect for a stratified weakly divergent coronal structure, at times $t = 200$ and $t = 600$. Comparing Fig.~\ref{contour}(a) and Fig.~\ref{contour}(b), we clearly see that as the Alfv\'en wave propagates along the $z$-axis, the wavefront is strongly dissipated around $x = 0$, where the transverse density gradient is strongest.

We note that the reduction in wave amplitude with height, due to enhanced phase mixing, results from two separate effects; firstly from the reduction in wavelength and the associated increased dissipation (as discussed above), and secondly from the area-divergence, or geometrical spreading, of the wave propagating in a divergent magnetic field. We also note that the characteristic transverse scale does not approach zero, even in strongly divergent coronal structures; as the transverse scale gets smaller, stronger transverse gradients build up until a balance is eventually reached with viscous dissipation. In the kinetic regime, the transverse scale can, as a result of strong phase mixing, be reduced down to the ion-cyclotron radius.

We see from Fig.~\ref{wavelengths}(a) that, in non-stratified coronal structures, increasing the divergence of the magnetic field, by lowering
$H_\mathrm{b}$, increases the rate at which the wavelength shortens with height. When $H_\mathrm{b} = 40$ the wavelength is reduced by approximately a factor of 4 from $\lambda_{||} \approx 5.5$ at $z = 0$, to $\lambda_{||} \approx 1.5$ at $z = 50$. From Eq.~(\ref{pdsgen1}), where $\Lambda \propto \omega^2 \propto 1/\lambda_{||}^2$, we see that a factor of 4 reduction in wavelength increases the dissipation rate by a factor of 16. In the stratified coronal structures of Fig.~\ref{wavelengths}(b) the picture is less straightforward; when $H_\mathrm{b} < 100$ the wavelength again reduced with height, but not to the same extent as in non-stratified coronal structures. When $H_\mathrm{b} = 40$ the wavelength is reduced by approximately a factor of 2 from $\lambda_{||} \approx 5.5$ at $z = 0$, to $\lambda_{||} \approx 2.5$ at $z = 50$. For $H_\mathrm{b} = 100$ the wavelength remained unchanged, corresponding to the special case where $H_\mathrm{\rho} = \frac{1}{2}H_\mathrm{b}$; see Eqs.~(\ref{pds3_l})--(\ref{pds3_lb}). This results from a constant Alfv\'en velocity with height where the reduction in Alfv\'en velocity, due to the divergence of the magnetic field, is exactly balanced by an increase in Alfv\'en velocity due to density stratification. This specific configuration does not exhibit the wavelength reduction associated with enhanced phase mixing, but it does still exhibit the effects of area-divergence. When $H_\mathrm{b} > 100$, the wavelength is seen to increase with height, due to the Alfv\'en velocity's inverse dependence on density, $V_\mathrm{A} \propto 1/\sqrt{\rho_0}$, which exponentially reduces with height as a result of stratification. In Fig.~\ref{wavelengths}(a), $H_\mathrm{b} = \infty$ is equivalent to a coronal structure permeated by a uniform magnetic field, and thus represents standard \citet{hp83} phase mixing. Comparing this to Fig.~\ref{wavelengths}(b), where $H_\mathrm{b} = \infty$ represents phase mixing in a stratified coronal structure, we see that density stratification works to increase the wavelength, and thus reduce the overall dissipation rate.

\subsection{Weakly divergent coronal structures}

In Figs.~\ref{amp_hpinf} and \ref{amp_hp50}, the numerically calculated Alfv\'en wave velocity amplitude, $V_\mathrm{y}$, is plotted as a function of height for both non-stratified and stratified weakly divergent coronal structures. These plots represent cross-sections of the wave's amplitude along $x = 0$, where the maximum dissipation occurs. We have also plotted our analytical solutions given by Eqs.~(\ref{pds4_l})--(\ref{pds4_lb}), as well as the corresponding equilibrium solutions of \citet{rnr98}. We have chosen to plot the incorrect original \citet{rnr98} solution to demonstrate the importance of our correction to the enhanced phase mixing solution of Sect.~3. Also note that we have only plotted $V_\mathrm{y}$ and not $B_\mathrm{y}$, since the behaviors of these two quantities are identical for non-stratified structures, and differ only slightly in stratified structures due to the amplification effect (see below).

Firstly from Figs.~\ref{amp_hpinf} and \ref{amp_hp50}, we note that lowering $H_\mathrm{b}$ increases the rate of dissipation (see Sect.~5.1). Secondly we see that in stratified structures only, there is an amplification of $V_\mathrm{y}$ at low heights, along with a corresponding reduction in $B_\mathrm{y}$ (not shown). This well known effect results from the wave amplitude's dependance on density; $V_\mathrm{y} \propto \rho_0^{-1/4}$ and $B_\mathrm{y} \propto \rho_0^{1/4} $ \citep[see, e.g.,][]{wg98,m01}. The reduction in density and subsequent increase in Alfv\'en velocity associated with stratification, therefore leads to the amplification of $V_\mathrm{y}$ seen in Fig.~\ref{amp_hp50}. We note that by using very high values of viscosity ($\nu \ge 10^{-3}$, corresponding to $\nu \ge 10^9$~m$^2$~s$^{-1}$), it is possible to remove this wave amplification, as viscous dissipation would dominate any amplification at low heights. Thirdly we see that, for a fixed magnetic scale height, $H_\mathrm{b}$, Alfv\'en waves dissipate slower in stratified structures than in non-stratified structures. This is due to the Alfv\'en velocity decreasing faster in non-stratified structures than in stratified structures. \citet{rnr98} suggested that at low heights, harmonic Alfv\'en waves dissipate according to the standard $\exp(-z^3)$ rate, since at these heights the reduction in wavelength required by enhanced phase mixing is not significant. This is confirmed by both Figs.~\ref{amp_hpinf} and \ref{amp_hp50} which show that, at low heights ($z \le 20$), the waves dissipate independently of $H_\mathrm{b}$. At larger heights ($z > 20$), the enhanced phase mixing mechanism begins to differentiate among the different equilibrium configurations.

\begin{figure}[t]
\includegraphics[width=\linewidth]{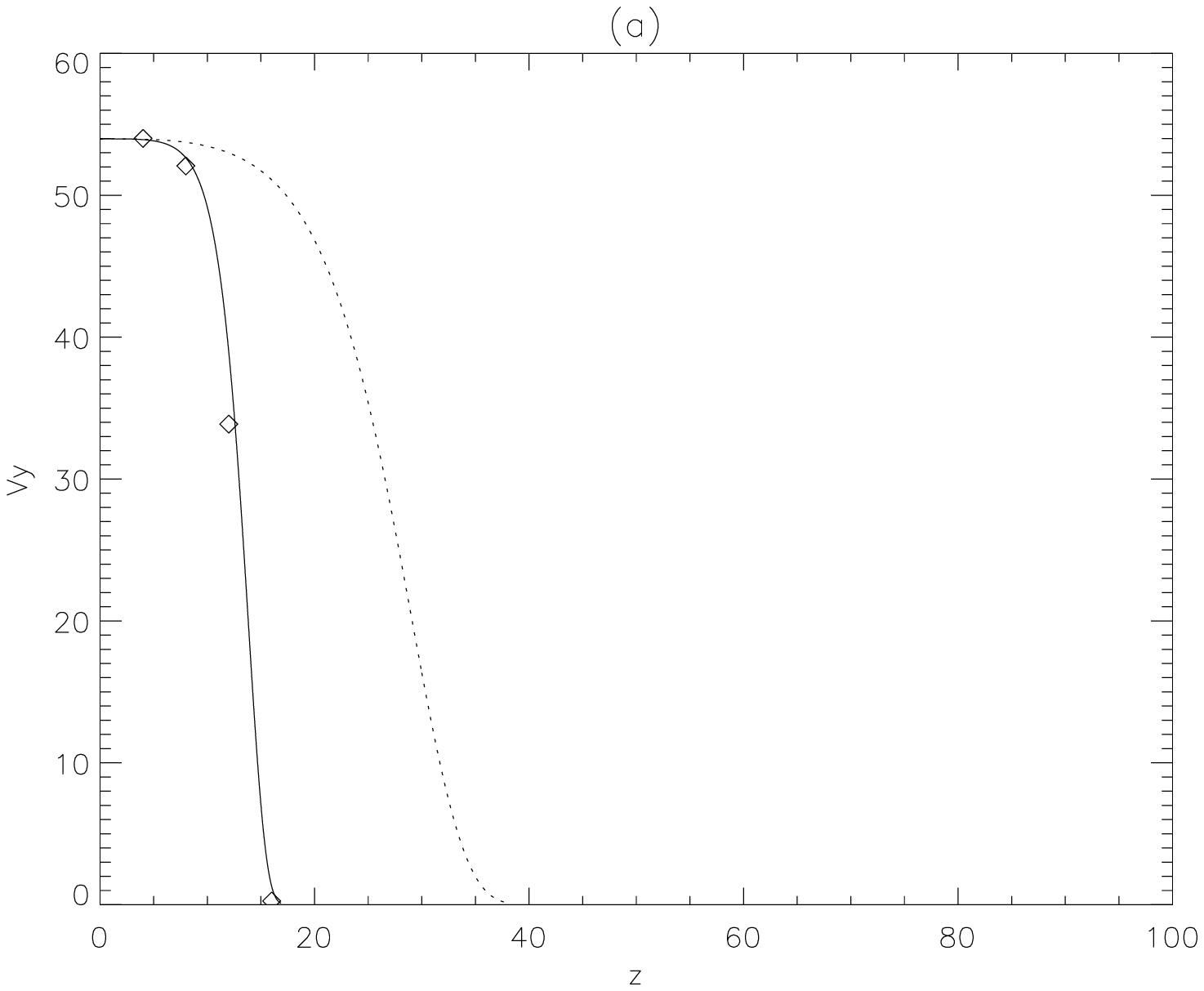}
\includegraphics[width=\linewidth]{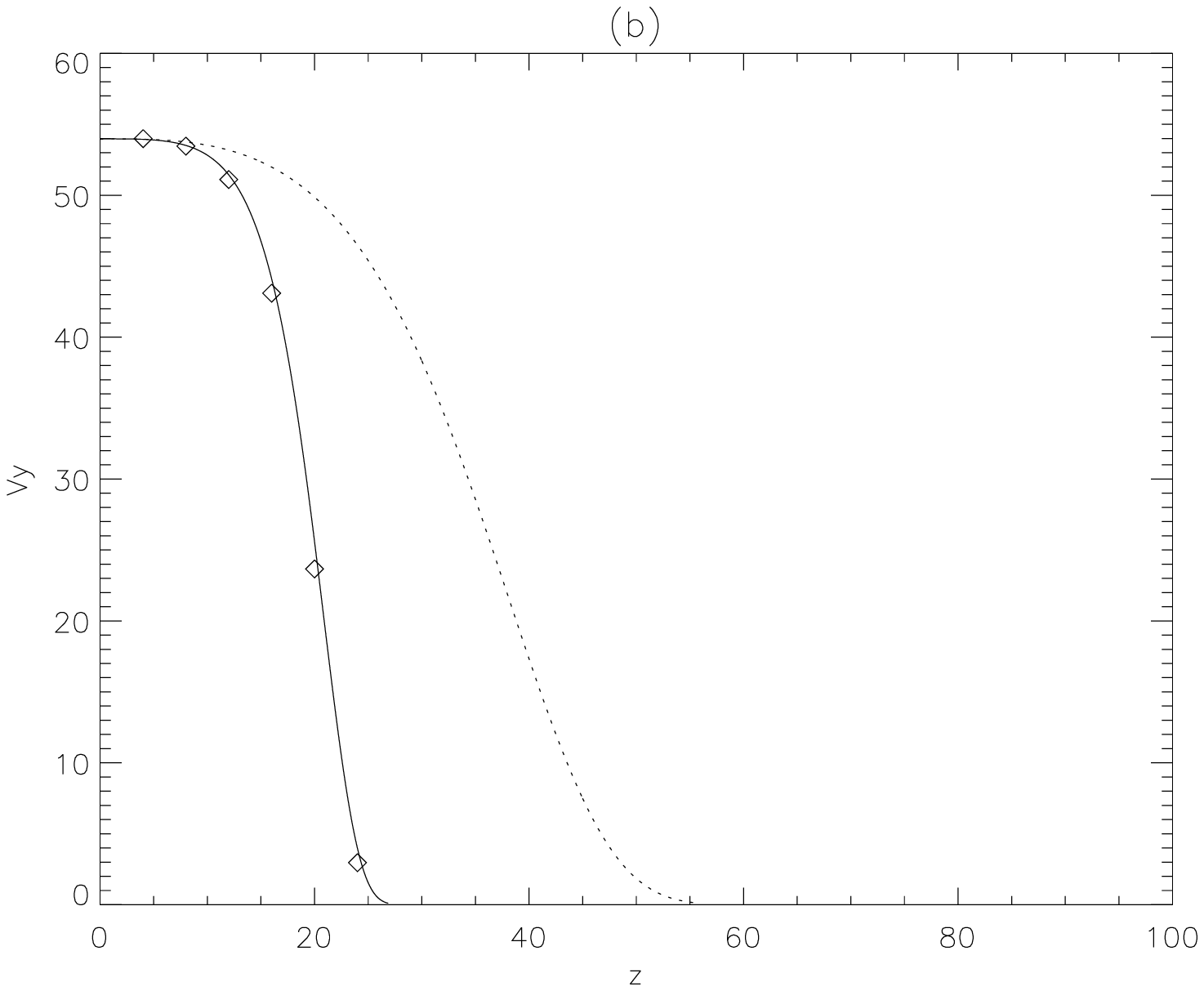}
\caption{Alfv\'en wave amplitude as a function of height along $x = 0$ for non-stratified, $H_\mathrm{\rho} = \infty$, with strongly divergent magnetic field: (a) $H_\mathrm{b} = 5$, (b) $H_\mathrm{b} = 10$. The solid and dotted lines shows calculations from the corrected and original \citet{rnr98} solution respectively, while the diamonds shows the numerical calculations. The dimensional units of amplitude, $V_\mathrm{y}$, and height, $z$, are respectively km~s$^{-1}$ and Mm.}\label{fig_vy_strong_inf}
\end{figure}
\begin{figure}[t]
\includegraphics[width=\linewidth]{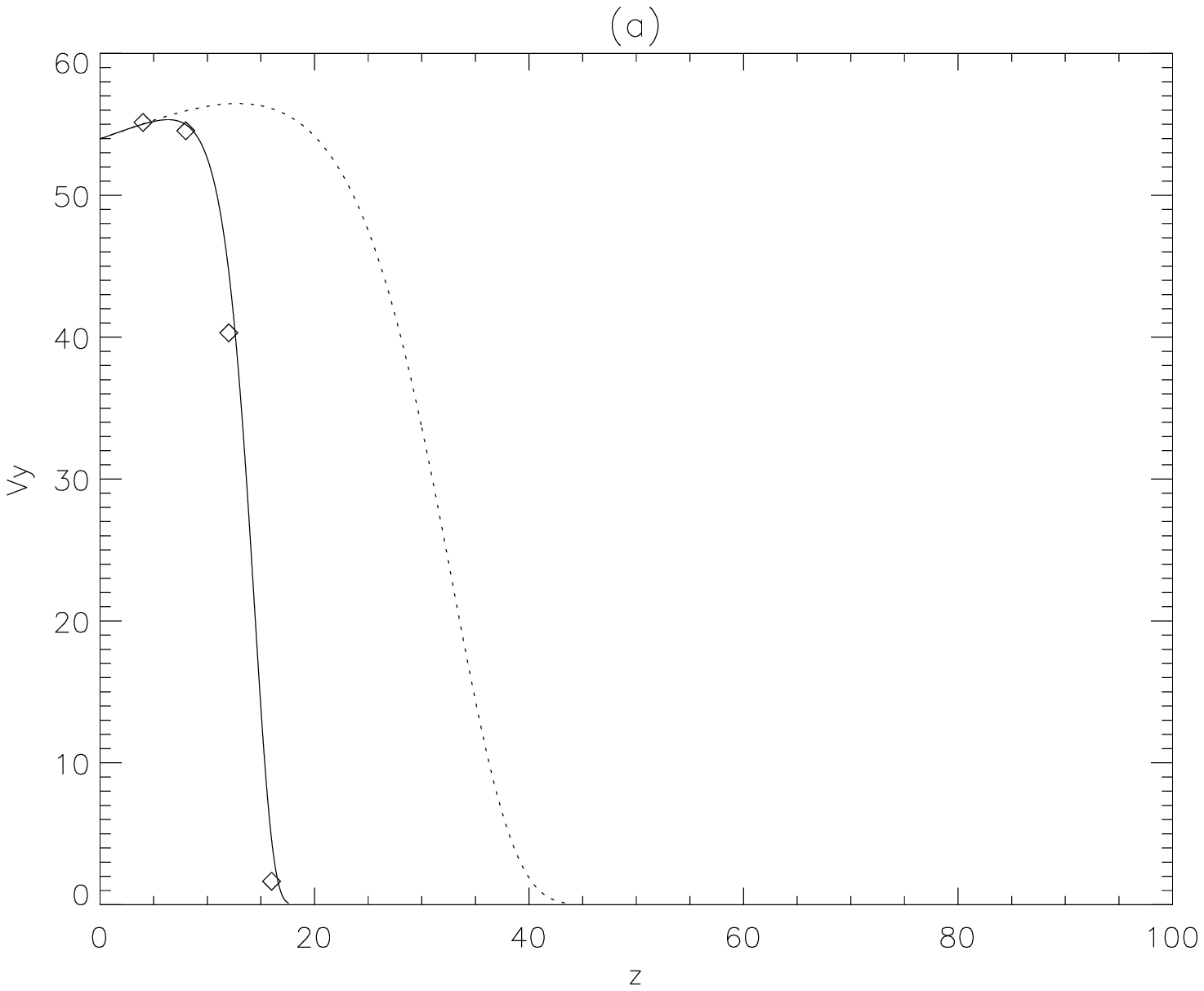}
\includegraphics[width=\linewidth]{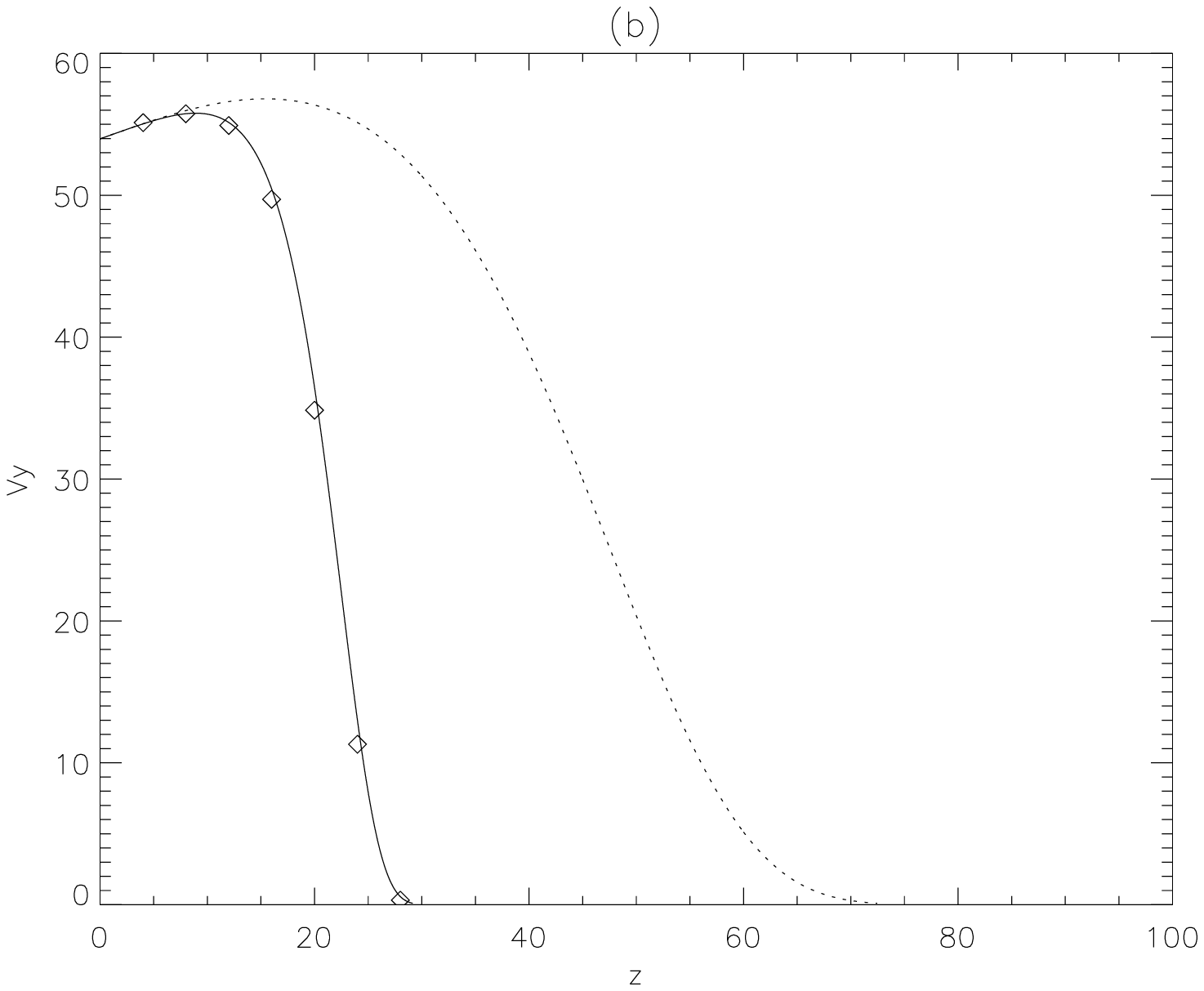}
\caption{Alfv\'en wave amplitude as a function of height along $x = 0$ for stratified, $H_\mathrm{\rho} = 50$, with strongly divergent magnetic field: (a) $H_\mathrm{b} = 5$, (b) $H_\mathrm{b} = 10$. The solid and dotted lines shows calculations from the corrected and original \citet{rnr98} solution respectively, while the diamonds shows the numerical calculations. The dimensional units of amplitude, $V_\mathrm{y}$, and height, $z$, are respectively km~s$^{-1}$ and Mm.}\label{fig_vy_strong_50}
\end{figure}

Next we compare our numerical calculations to our analytical solutions given by Eqs.~(\ref{pds4_l})--(\ref{pds4_lb}), as well as to the corresponding equilibrium solutions of \citet{rnr98}. In Figs.~\ref{amp_hpinf}(d) and \ref{amp_hp50}(d), representing phase mixing in a uniform magnetic field, we see strong agreement between the numerical model and both analytical solutions. As the magnetic scale height is lowered from $H_\mathrm{b} = \infty$ to $H_\mathrm{b} = 40$, the analytical solutions of \citet{rnr98} begin to diverge from the numerical model results, whereas our analytical solutions continue to show good agreement over a range of magnetic scale heights. This is most apparent in the stratified numerical calculations, Fig.~\ref{amp_hp50}, where the lower overall dissipation rates enable the difference between standard and enhanced phase mixing to be more clearly seen. Notably when $H_\mathrm{b} = 40$ for the stratified numerical model, Fig.~\ref{amp_hp50}(a), the Alfv\'en wave is almost completely dissipated by $z \approx 60$; half the height at which the \citet{rnr98} analytical solution predicts and within one density scale height, $H_\mathrm{\rho}$.

\subsection{Strongly divergent coronal structures}

We now consider coronal structures with strongly divergent magnetic fields, where the magnetic scale height is of the order of the characteristic transverse length scale, ($x_0/H_\mathrm{b} \approx 1$). This clearly violates the first assumption, $x_0/H \ll 1$, used to derive the general analytical solution (See Sect.~3). We would therefore expect to see a large difference between the numerical and analytical solution.

In Figs.~\ref{fig_vy_strong_inf} and ~\ref{fig_vy_strong_50}, we plot the numerically calculated Alfv\'en wave amplitude, $V_\mathrm{y}$, along $x = 0$ for respectively, non-stratified and stratified strongly divergent coronal structures. As in Sect.~5.2 we also plotted our analytical solutions given by Eqs.~(\ref{pds4_l})--(\ref{pds4_lb}), as well as the corresponding equilibrium analytical solutions of \citet{rnr98}. Comparing the two plots, we see that the effects of stratification are less pronounced for $H_\mathrm{b} = 5$, than for $H_\mathrm{b} = 10$. This occurs as the wave amplitude's dependence on magnetic scale height, $\exp(-\exp(z/H_\mathrm{b}))$, dominates over the wave amplitude's dependance on viscosity and frequency, $\exp(-\nu \omega^2)$, in highly divergent magnetic fields. We therefore focus on Fig.~\ref{fig_vy_strong_50}, where once again we see that decreasing $H_\mathrm{b}$ increases the dissipation rate as expected. We see that, when $H_\mathrm{b} = 5$, we can fully dissipate Alfv\'en waves with $f = 0.1$ within $z \approx 20$, which is over six times lower than would occur in the standard \citet{hp83} phase mixing case; Fig.~\ref{amp_hpinf}(d). This is also half the density scale height $H_\mathrm{\rho}$, as well as half the height that the \citet{rnr98} analytical solutions predict, which as in the weakly divergent coronal structures, differs strongly from both our analytical and numerical calculations.

Note that the amplification of $V_\mathrm{y}$ due to density stratification (see Sect.~5.2), is less pronounced than in the coronal structures with weakly divergent magnetic field seen in Fig.~\ref{amp_hp50}. This is because the strongly divergent magnetic field very quickly generates strong dissipation at heights much lower than in the weakly divergent case, which is then able to overpower the stratification effect. Surprisingly the numerical results are still in good agreement with the corrected analytical solution along $x = 0$, even when the magnetic scale height is lowered to $H_\mathrm{b} = 5$ ($x_0/H_\mathrm{b} \approx 0.2)$. It is interesting to note that standard phase mixing modeled by PIC (Particle in Cell) codes in the kinetic regime, gives dissipation rates in strong agreement with the \citet{hp83} MHD dissipation rates \citep[see][]{tss05b}. The effective resistivity seen in this paper, as well as in \citet{t06b}, were also found to be many orders higher than the classical Braginskii value. This result along with the demonstration that two physical descriptions, valid on completely different scales, can lead to the same dissipation rates, strongly supports the use of MHD and anomalous viscosity in this study.

\subsection{Viscous heating}

We now discuss the viscous heating generated by the enhanced phase mixing of Alfv\'en waves in divergent coronal structures. This heating results from the phase mixing mechanism converting the large scale wave energy to the small scale thermal energy of the plasma, via viscous (or resistive) dissipation. In this section, we calculate the generated viscous heating power, $E_\mathrm{H}$, and viscous heating flux, $F_\mathrm{H}$, from the numerical calculations of Sects.~5.2 and 5.3.

\begin{figure}[t]
\includegraphics[width=\linewidth]{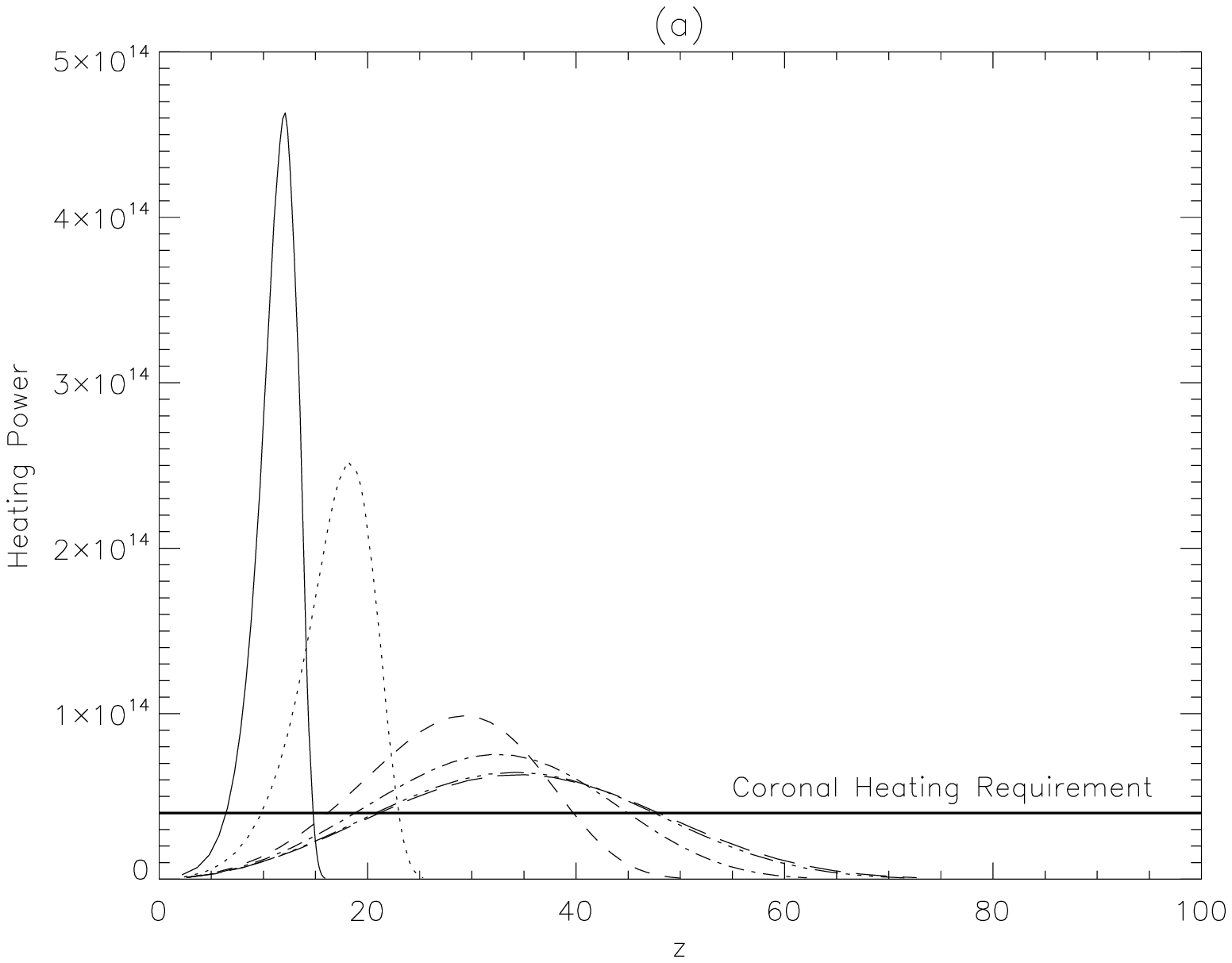}
\includegraphics[width=\linewidth]{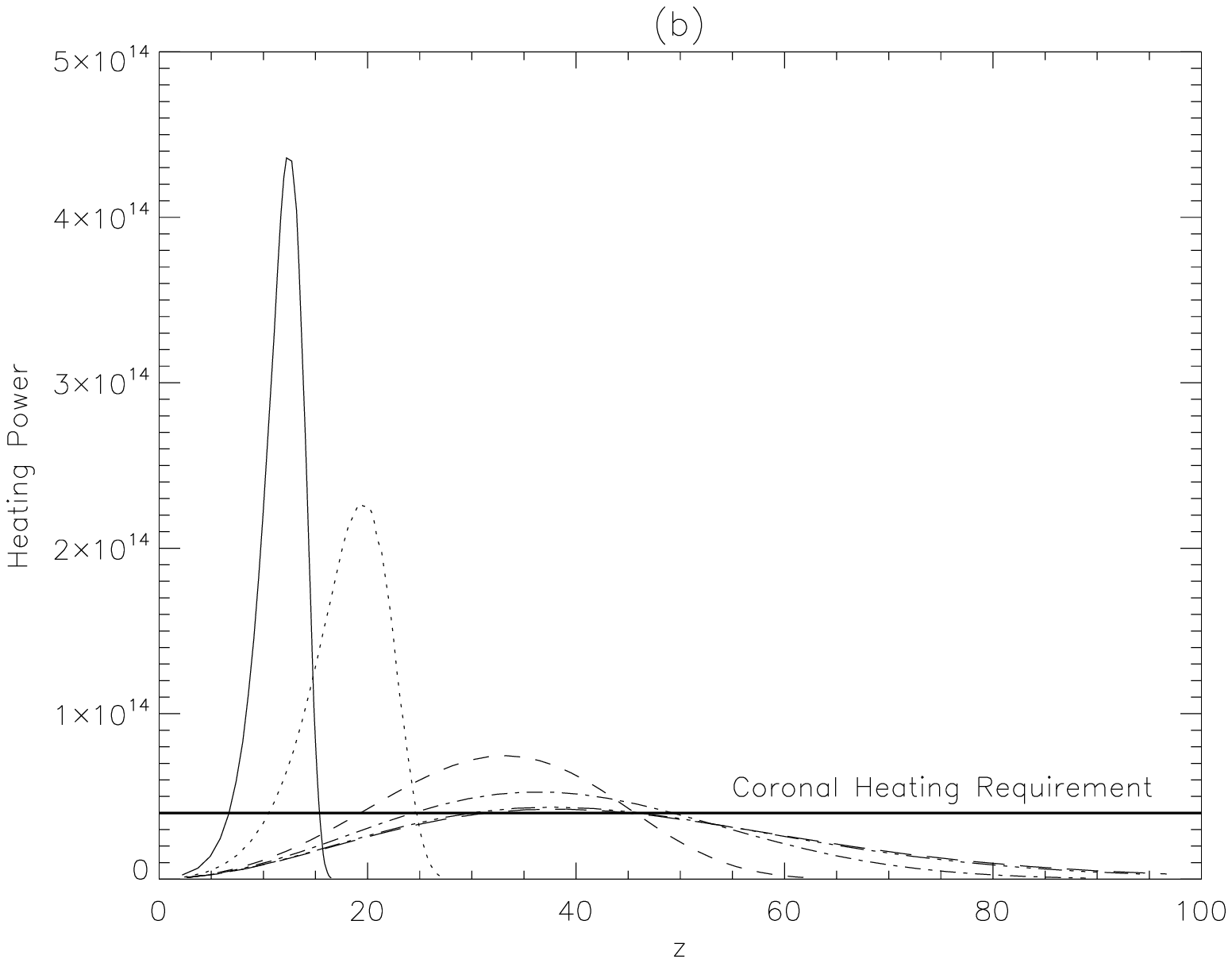}
\caption{Heating power along a flux tube for (a) non-stratified, $H_\mathrm{\rho} = \infty$, and (b) stratified, $H_\mathrm{\rho} = 50$, divergent coronal structures. Solid, dotted, dashed, dot-dashed, triple-dot-dashed and long dashed lines correspond respectively to $H_\mathrm{b} = 5$, $H_\mathrm{b} = 10$, $H_\mathrm{b} = 40$, $H_\mathrm{b} = 100$, $H_\mathrm{b} = 700$ and $H_\mathrm{b} = \infty$. The dimensional units of heating power, $E_\mathrm{H}$, are $\times 10^{-18}$~J~m$^{-3}$~s$^{-1}$, while those for $H_\mathrm{b}$ and $z$ are Mm.}
\label{heat_rate}
\end{figure}

\begin{figure}[t]
\includegraphics[width=\linewidth]{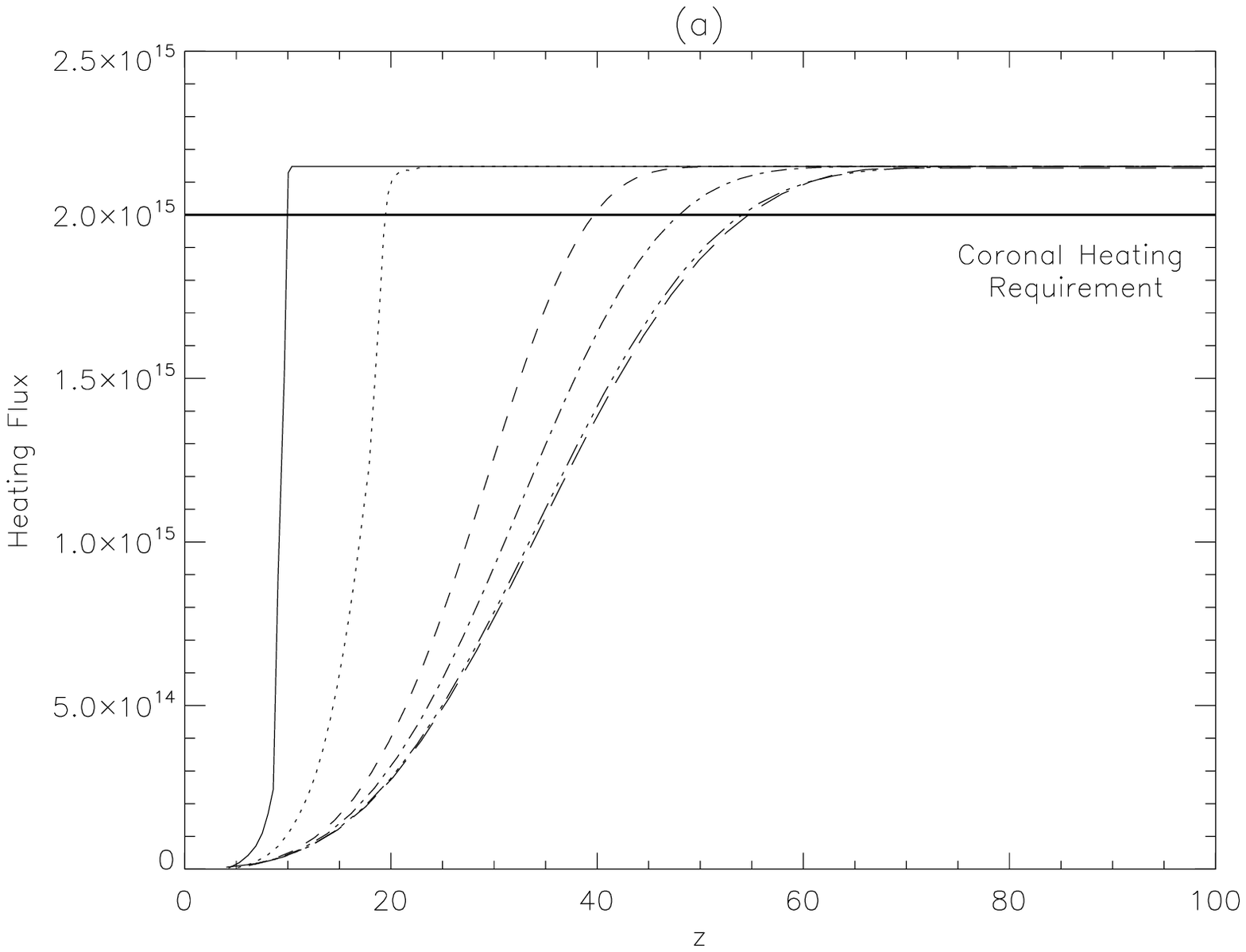}
\includegraphics[width=\linewidth]{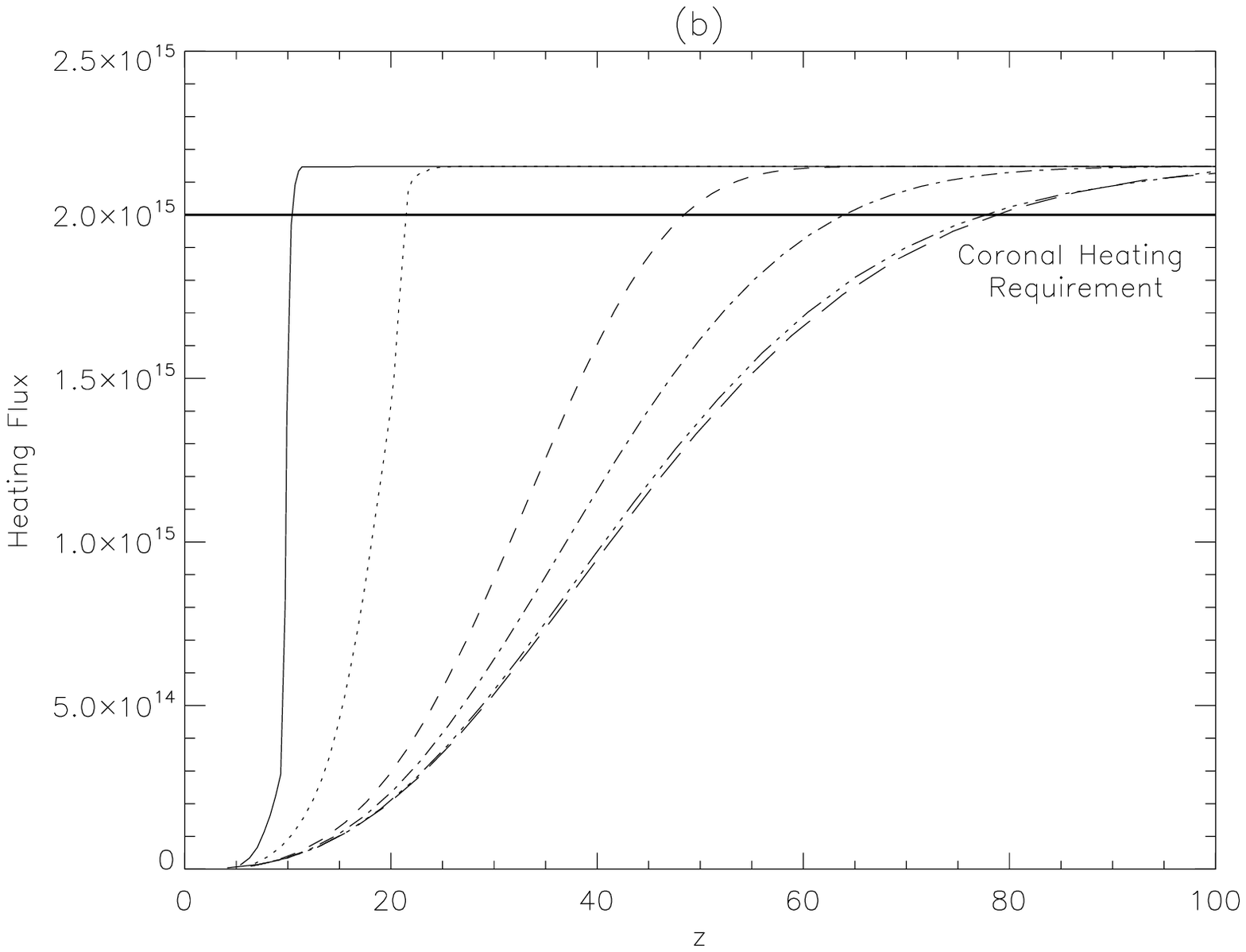}
\caption{Heating flux along a flux tube for (a) non-stratified, $H_\mathrm{\rho} = \infty$, and (b) stratified, $H_\mathrm{\rho} = 50$, divergent coronal structures. Solid, dotted, dashed, dot-dashed, triple-dot-dashed and long dashed lines correspond respectively to $H_\mathrm{b} = 5$, $H_\mathrm{b} = 10$, $H_\mathrm{b} = 40$, $H_\mathrm{b} = 100$, $H_\mathrm{b} = 700$ and $H_\mathrm{b} = \infty$. The dimensional units of heating flux, $F_\mathrm{H}$, are $\times 10^{-12}$~J~m$^{-2}$~s$^{-1}$, while those for $H_\mathrm{b}$ and $z$ are Mm.}
\label{heat_flux}
\end{figure}

The viscous heating power describes the location and magnitude of the energy dissipation, and is calculated from
\begin{equation}
E_\mathrm{H} \left( {x,z} \right) = \nu \rho_0 \left( {\nabla  \times {\vec{V}}} \right)^2 e^{z/H_\mathrm{b}}.
\label{heat_rate_eq}
\end{equation}
The viscous heating flux describes the flow of energy with the upward propagating waves, and is derived from the Poynting vector
\begin{equation}
\vec{S} = \frac{\vec{E} \times \vec{B}_1}{\mu_0},
\label{poynting_flux_eq}
\end{equation}
where subscripts 0 and 1 refer respectively to equilibrium and perturbation quantities (except for $\mu_0$ which is the magnetic permeability), and $\vec{E}$ is the electric field given by $\vec{E} = - \vec{V}_1 \times \vec{B}_0$. Substituting this into Eq.~(\ref{poynting_flux_eq}) gives an alternative form for the wave Poynting vector
\begin{equation}
\vec{S} = \frac{ \vec{B}_1 \times (\vec{V}_1 \times \vec{B}_0)}{\mu_0},
\label{poynting_flux_eq2}
\end{equation}
from which we derive the wave Poynting flux, $\langle \vec{S} \rangle = |B_\mathrm{y} V_\mathrm{y}| B_0 / \mu_0$, and finally heating flux, $F_\mathrm{H} = F_\mathrm{0}  - \langle \vec{S} \rangle $, as a function of height
\begin{equation}
F_\mathrm{H} \left( {x,z} \right) = F_\mathrm{0} - \frac{|B_\mathrm{y} V_\mathrm{y}| B_0}{\mu_0} e^{z/H_\mathrm{b}},
\label{heat_flux_eq}
\end{equation}
where $F_\mathrm{0} = A_0^2 B_{00}^2 \bar{V}_{\mathrm{A}}/\mu_0$ is the wave Poynting flux at the coordinate origin, $x = z = 0$.
Note that the heating rate and heating flux oscillate in time at twice the wave frequency, which ideally would be removed by time averaging Eqs.~(\ref{heat_rate_eq}) and (\ref{heat_flux_eq}). Instead, for numerical reasons, we have taken the wave envelope, scaled by $0.5$, to be equivalent to time averaging when plotting Figs.~\ref{heat_rate} and \ref{heat_flux}. The $\exp(z/H_\mathrm{b})$ terms seen in Eqs.~(\ref{heat_rate_eq}) and (\ref{heat_flux_eq}), compensate for the area divergence of each flux tube, so that the effects of standard and enhanced phase mixing on wave dissipation, can be directly compared.

In Figs.~\ref{heat_rate} and \ref{heat_flux}, we show how the viscous heating power and viscous heating flux along a flux tube, depend on the magnetic scale height. We begin by comparing Figs.~\ref{heat_rate}(a) and ~\ref{heat_rate}(b), where we see that the effects of density stratification on the magnitude and location of peak viscous heating power, is minor. There is a slight reduction in the magnitude of peak heating power and lowering of the height at which this occurs, but these effects become increasingly insignificant as the magnetic field becomes more divergent ($H_\mathrm{b}$ is decreased). Therefore we focus on the more realistic case of Fig.~\ref{heat_rate}(b), where the effects of enhanced phase mixing are clearly apparent; increasing the magnetic field divergence, by reducing $H_\mathrm{b}$, causes; (i) the magnitude of heating power to increase, and (ii) the location of the heating power peak to be lowered. Indeed, comparing a strongly divergent magnetic field, $H_\mathrm{b} = 5$, to a uniform magnetic field, $H_\mathrm{b} = \infty$, we see that the viscous heating power is increased by a factor of ten, from $E_\mathrm{H} \approx 4.2 \times 10^{13}$ to $E_\mathrm{H} \approx 4.4 \times 10^{14}$. This is approximately ten times the corresponding heating power requirement for an active region; $E_\mathrm{H} \approx 4.0 \times 10^{13}$. We also see that the location of peak heating is lowered by a factor of three, from $z \approx 40$ to $z \approx 13$.

Next we compare Figs.~\ref{heat_flux}(a) and \ref{heat_flux}(b), where we see that density stratification affects weakly divergent coronal structures far more than their strongly divergent counterparts. This occurs because, in strongly divergent coronal structures, the wavelength reducing effects of enhanced phase mixing dominate over the wavelength increasing effects of density stratification (See Sect.~5.1). We again focus on the more realistic stratified case, Fig.~\ref{heat_flux}(b), where firstly we note that the maximum viscous heating flux carried by the Alfv\'en waves along a flux tube, is $F_\mathrm{H} \approx 2.1 \times 10^{15}$. Given that the heating flux requirement for an active region is $F_\mathrm{H} \approx 2.0 \times 10^{15}$, this agrees well with our expectation that Alfv\'en waves carry sufficient energy to heat the corona. Now comparing a strongly divergent magnetic field, $H_\mathrm{b} = 5$, to a uniform magnetic field, $H_\mathrm{b} = \infty$, we see that the height at which the Alfv\'en waves have dissipated $95\%$ of their heating flux, which we take to be representative of the heating length scale, $L_\mathrm{H}$, is lowered by a factor of six; from $L_\mathrm{H} \approx 90$ to $L_\mathrm{H} \approx 15$. This follows from Sect.~5.3 where we showed that Alfv\'en waves could be fully dissipated over six times lower in a strongly divergent field, than in a uniform magnetic field.

\subsection{Heating length scale}

\begin{figure*}[t]
\includegraphics[width=0.5\linewidth]{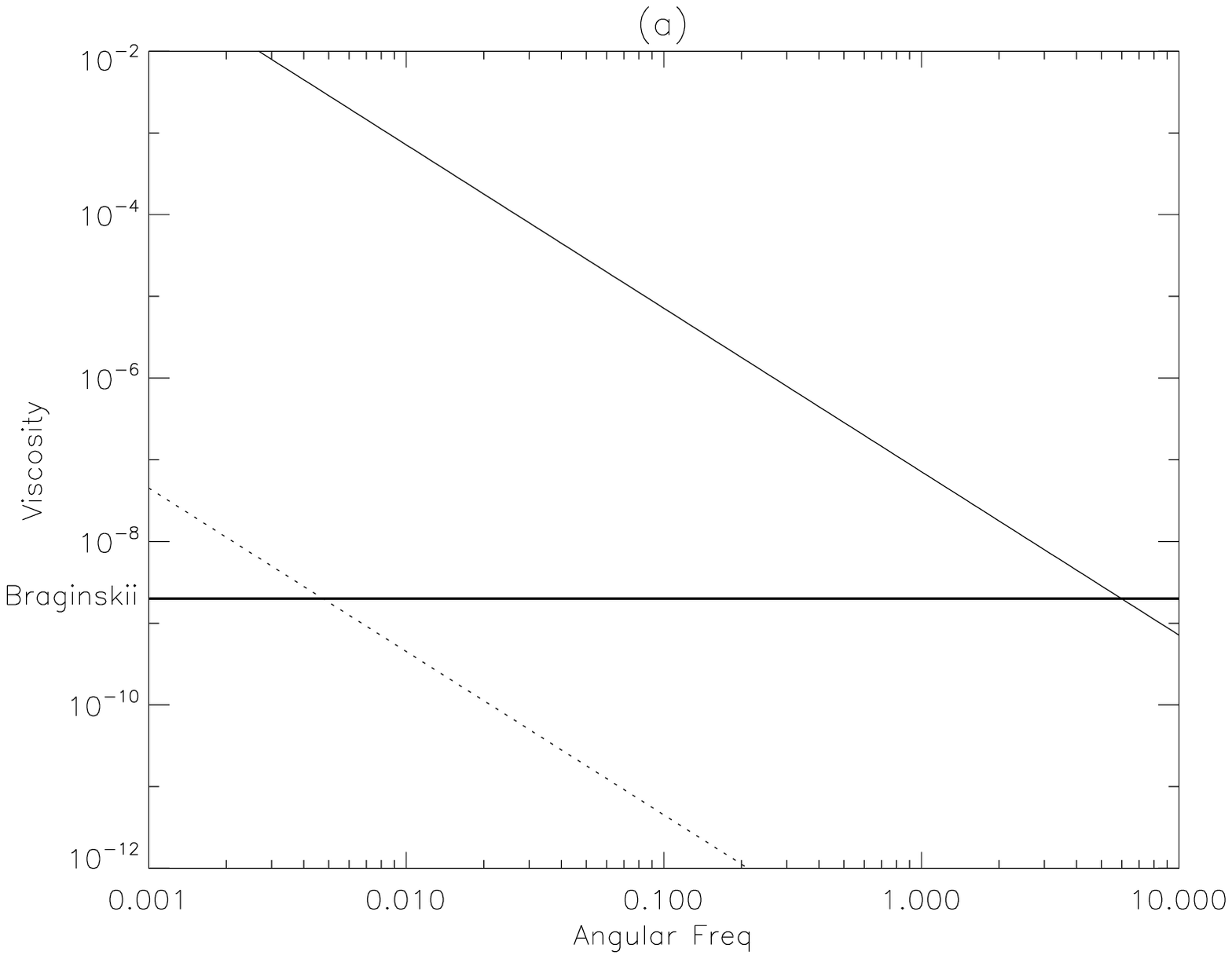}
\includegraphics[width=0.5\linewidth]{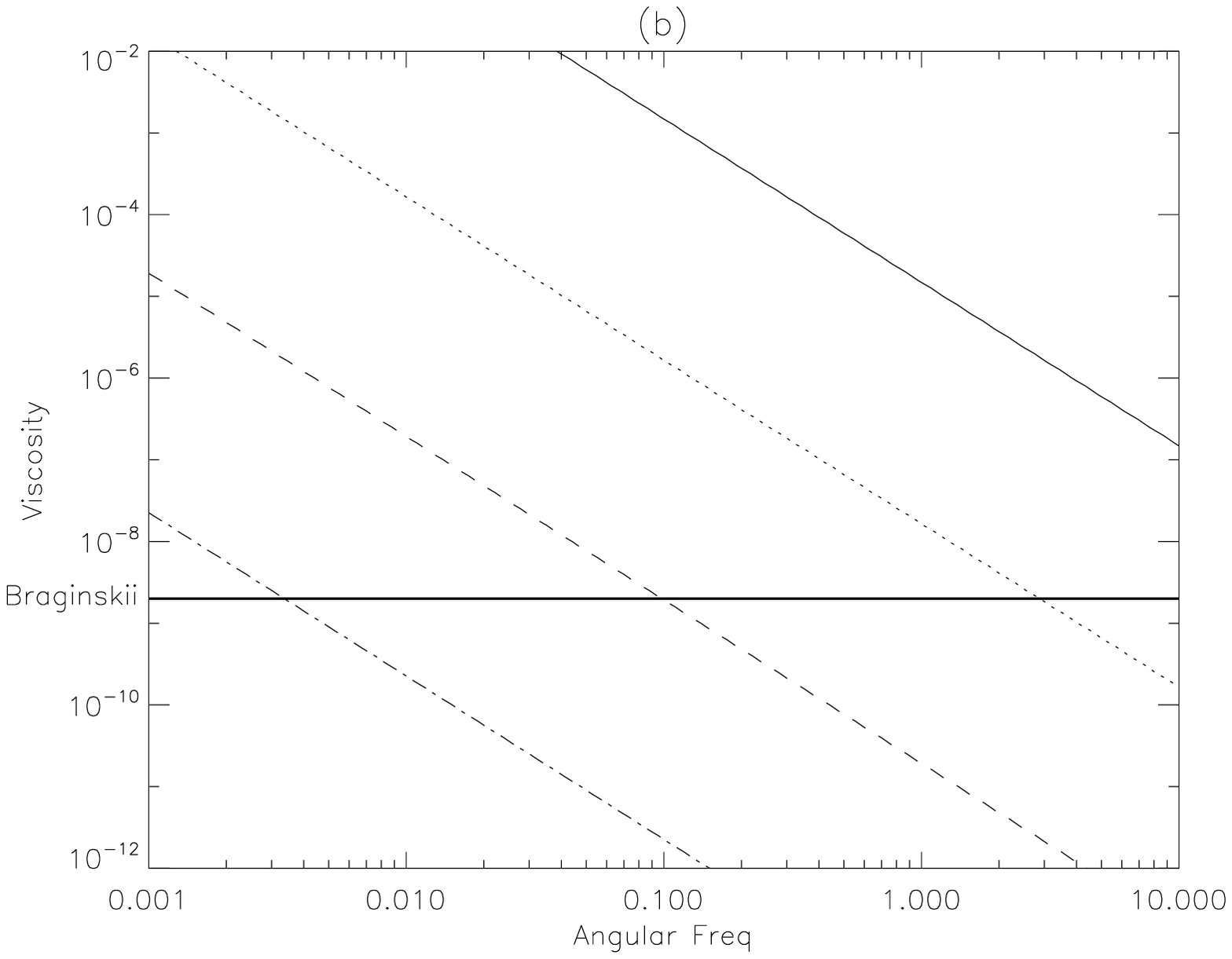}
\includegraphics[width=0.5\linewidth]{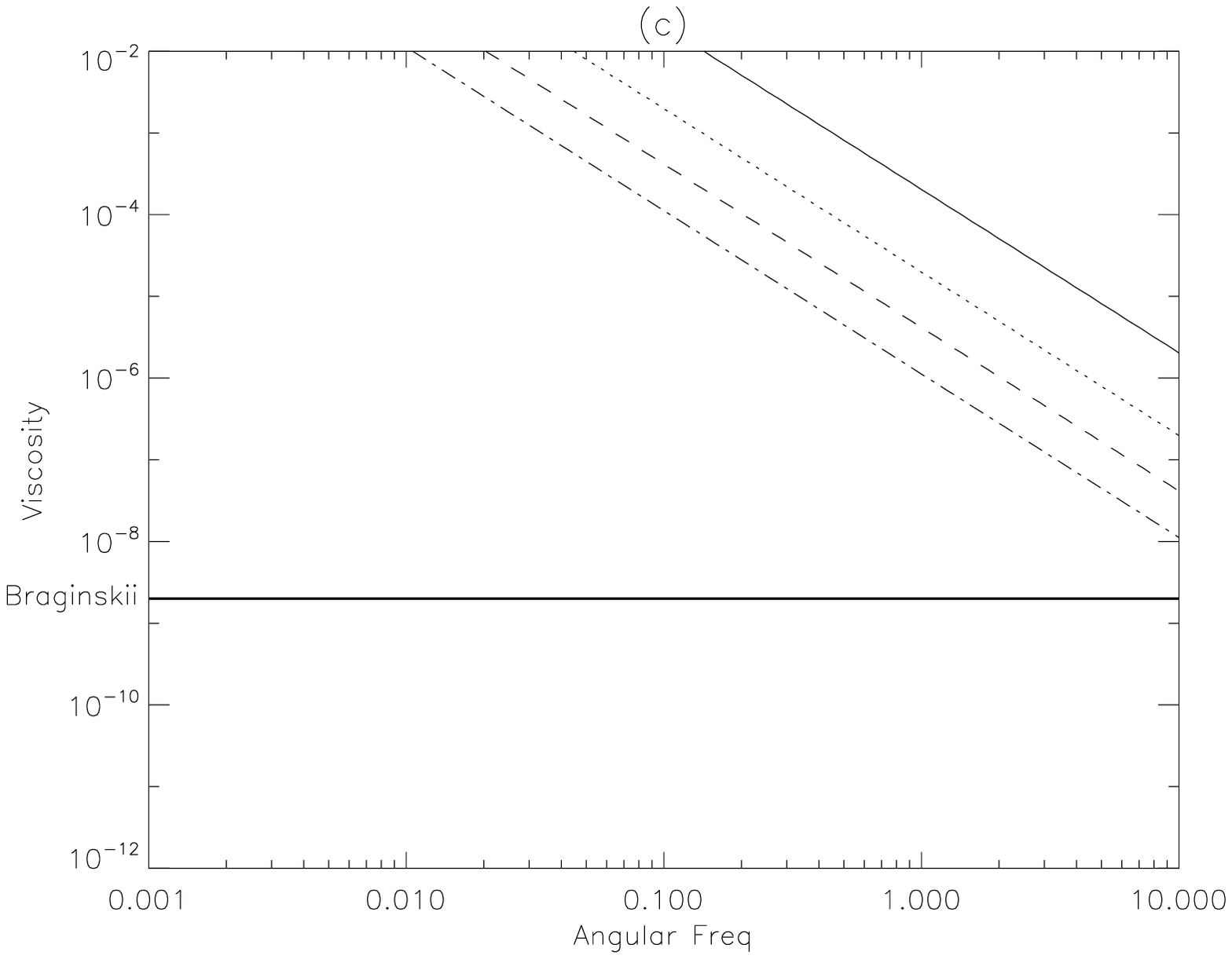}
\includegraphics[width=0.5\linewidth]{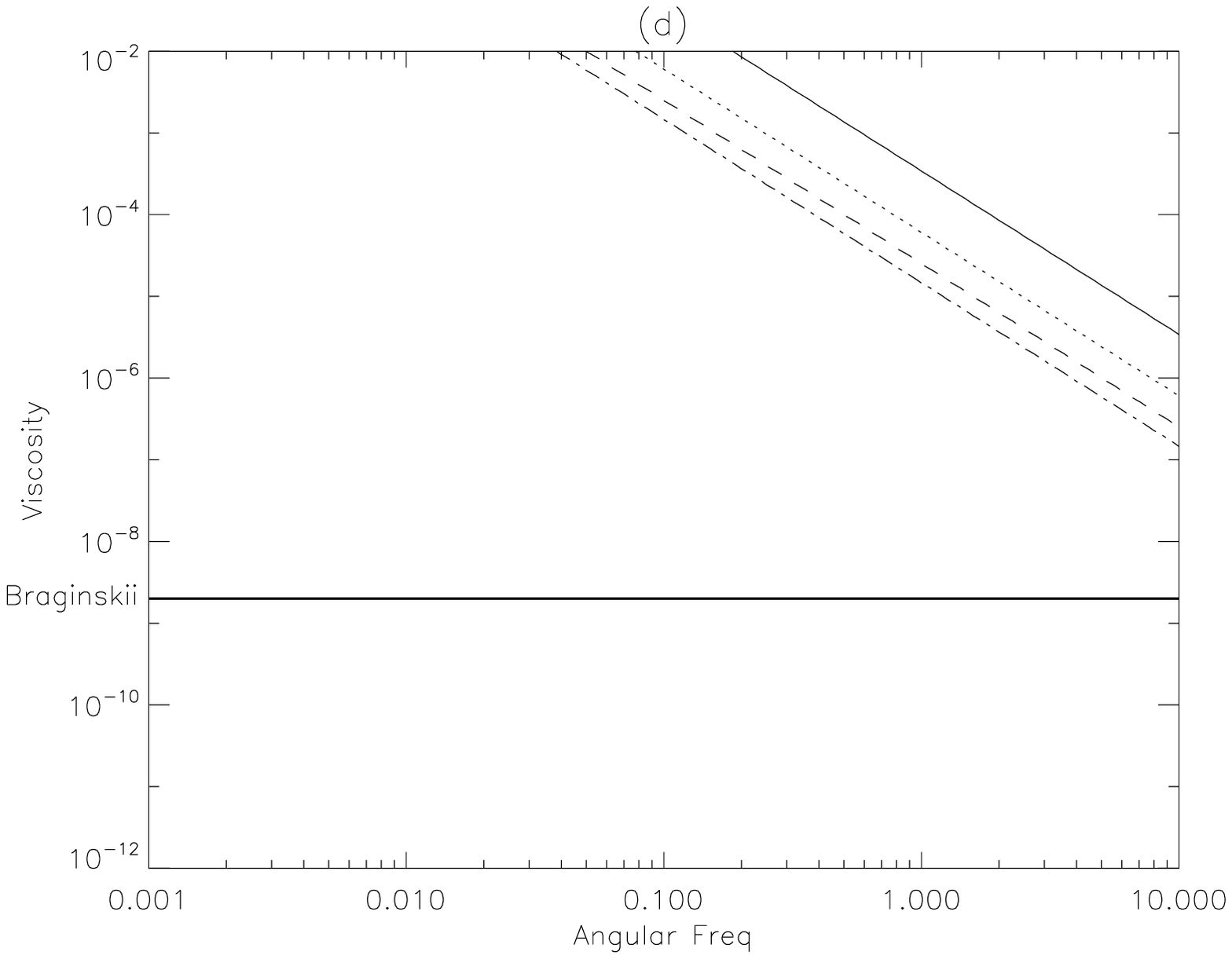}
\caption{Heating length scale as a function of driving frequency, $\omega$, and shear viscosity, $\nu$, where (a) $H_\mathrm{b} = 5$, (b) $H_\mathrm{b} = 10$, (c) $H_\mathrm{b} = 40$ and (d) $H_\mathrm{b} = \infty$. Solid, dotted, dashed and dot-dashed lines correspond to $L_\mathrm{H} = 25$, $L_\mathrm{H} = 50$, $L_\mathrm{H} = 75$ and $L_\mathrm{H} = 100$ respectively ($H_\rho = 50$). The dimensional units of $L_\mathrm{H}$ and $H_\mathrm{b}$ are Mm, $\omega$ are radians $s^{-1}$ and $\nu$ are $\times 10^{12}$~m$^2$~s$^{-1}$.}
\label{fig_omega_nu}
\end{figure*}

In the previous section we demonstrated that for the parameters chosen in this study, which were based on observational evidence, Alfv\'en waves contain sufficient energy to fulfill the coronal heating requirement. The only question is therefore at what height is this fulfillment achieved? To answer this, we consider the effects of altering the driving frequency, $\omega$, and the shear viscosity, $\nu$, on the heating length scale, $L_\mathrm{H}$, generated by the enhanced phase mixing of Alfv\'en waves propagating in divergent and stratified, $H_\mathrm{\rho} = 50$, coronal structures of varying magnetic field divergence. We use Eqs.~(\ref{pds4_l}) and (\ref{pds4_lb}), from the analytical solutions of Sect.~3, which were found to be in strong agreement with the numerical calculations of Sects.~5.2 and 5.3, to perform a parametric study of the variation in heating length scale with driving frequency and shear viscosity. In the previous section the heating length scale, $L_\mathrm{H}$, was defined to be the height at which $95\%$ of the Alfv\'en wave Poynting flux, along a flux tube, had been dissipated. For an active region, this heating length scale is also required to be within a density scale height, which in our model corresponds to $L_\mathrm{H} \leq 50$, to agree with an assertion by \citet{ana00} that the heating scale height is less than the density scale height.

In Fig.~\ref{fig_omega_nu}, we plot the heating length scale as a function of $\omega$ and $\nu$, for magnetic scale heights, $H_\mathrm{b} = 5, 10, 40, \infty$. Firstly, from the standard phase mixing case shown in Fig.~\ref{fig_omega_nu}(d), we see that to generate the required heating length scale of $L_\mathrm{H} \leq 50$ using observable, $\omega \approx 0.01$, Alfv\'en waves, requires an anomalous viscosity 8 orders of magnitude higher than classical Braginskii viscosity, $\nu \approx 2 \times 10^{-9}$. This demonstrates the main problem with standard phase mixing in the corona; we need to use extremely high values of shear viscosity, to dissipate Alfv\'en waves low enough in the corona to contribute to coronal heating. Secondly, we see that even for high frequency, $\omega \ge 1.0$, Alfv\'en waves, we still require the viscosity to be 3 to 5 orders of magnitude higher than the Braginskii value. Clearly, without the combination of anomalous viscosity and high frequency Alfv\'en waves, standard phase mixing is not a viable coronal heating mechanism.

In Fig.~\ref{fig_omega_nu}(c), we see that the enhanced phase mixing of Alfv\'en waves in weakly divergent magnetic fields, $H_\mathrm{b} = 40$, does not significantly alter the heating length scale. Whereas in Figs.~\ref{fig_omega_nu}(a) and \ref{fig_omega_nu}(b), which represent the enhanced phase mixing of Alfv\'en waves in strongly divergent magnetic fields, $H_\mathrm{b} \le 10$, the heating length scale is significantly reduced. For $H_\mathrm{b} = 5$, we see that the enhanced phase mixing of observable, $\omega \approx 0.01$, Alfv\'en waves generates the required heating length scales, $L_\mathrm{H} \le 50$, using classical Braginskii viscosity. As in Sect.~5.3, this occurs because in highly divergent magnetic fields, the wave amplitude dependence on magnetic scale height, $\exp(-\exp(z/H_\mathrm{b}))$, dominates over the wave amplitude dependance on viscosity and frequency, $\exp(-\nu \omega^2)$. Comparing Figs.~\ref{fig_omega_nu}(a) and \ref{fig_omega_nu}(d), we see that in standard phase mixing, small changes of $\omega$ or $\nu$ have large effects on the heating length scale, whereas the opposite is true for enhanced phase mixing in a strongly divergent magnetic field. We also see that, for a given wave frequency, to generate a heating length scale of $L_\mathrm{H} \le 50$ using enhanced phase mixing, requires a shear viscosity eight orders of magnitude lower than standard phase mixing. Finally, for $\omega \approx 0.6$ and $\nu = 5 \times 10^{-5}$, corresponding to the values used in this study, the magnetic scale height must be $H_\mathrm{b} \le 40$ for the heating length scale to be $L_\mathrm{H} \le 50$, which is in agreement with the corresponding numerical calculations of Sect.~5.4.

Note that we must exercise caution when considering low frequency Alfv\'en waves, since the derivation of the analytical solution of Sect.~3 assumes the wavelength is smaller than or of order the transverse scale, $x_0$. To demonstrate the accuracy of Fig.~\ref{fig_omega_nu}, we therefore conducted numerical calculations for the low frequency case, where $\omega \approx 0.01$ and $\nu = 2 \times 10^{-9}$ for $H_\mathrm{b} = 5$, $H_\mathrm{\rho} = 50$. We found the calculated heating length scale, $L_\mathrm{H} \approx 50$, to be in good agreement with the results seen in Fig.~\ref{fig_omega_nu}.

\section{Discussion and Conclusions}

We analytically and numerically studied the phase mixing of Alfv\'en waves propagating in weakly divergent, $H_\mathrm{b} = 40,100,700,\infty$~Mm, and strongly divergent, $H_\mathrm{b} = 5,10$~Mm, stratified, $H_\mathrm{\rho} = 50$~Mm, coronal structures. These numerical calculations were used to validate our analytical solution, which was obtained by correcting an error in the general analytical solution of \citet{rnr98}. For convenience, in this section we revert to using dimensional units.

We began in Sect.~5.1 by showing that density stratification and magnetic field divergence are two opposing factors affecting the wavelength;
stratification works to increase the Alfv\'en velocity while divergence works to decrease it. An increasing Alfv\'en velocity increases the
wavelength leading to larger transverse scales, and thus reduces the wave dissipation rates. Conversely a decreasing Alfv\'en velocity shortens
the wavelength leading to smaller transverse scales, and thus increases the wave dissipation rates. Note that in both cases, we assume that the angular wave frequency, $\omega$, remains fixed. We used this to define the concept of enhanced phase mixing as; phase mixing occurring in divergent magnetic fields, which results in stronger dissipation due to a reduction in the wavelength. We found that in stratified coronal structures, enhanced phase mixing occurs only when the magnetic scale height is less than twice the density scale height, $H_\mathrm{b}/H_\mathrm{\rho} < 2$. Therefore the enhanced phase mixing of Alfv\'en waves in the corona will only occur when $H_\mathrm{b} < 100$~Mm, given that the density scale height is $H_\mathrm{\rho} = 50$~Mm. This also means that enhanced phase mixing does not occur in typical coronal plumes where $H_\mathrm{\rho} \ll H_\mathrm{b}$, but is instead limited to the density boundaries of highly divergent coronal structures, e.g. coronal loops and arcades.

In Sect.~5.2 we compared our numerical calculations to our corrected analytical solution, Eqs.~(\ref{pdsgen2})--(\ref{pdsgen1}), as well as to the previous analytical solution of \citet{rnr98}. For very weakly divergent magnetic fields, corresponding to coronal plumes where $H_\mathrm{b} \ge 700$~Mm, our analytical and numerical calculations approximately agree with those of \citet{rnr98}. Their conclusion that enhanced phase mixing is not an important dissipation mechanism in coronal plumes (a very weakly divergent magnetic field), is valid. However due to a derivation error (see Sect.~3), the dissipation rates given by the \citet{rnr98} analytical solution show very small variation with $H_\mathrm{b}$, implying that magnetic field divergence has little effect on the dissipation of Alfv\'en waves. The numerical calculations we conducted in conjunction with our analytical solution show conclusively that this is not the case. Indeed for weakly divergent, $H_\mathrm{b} = 40$~Mm, stratified, $H_\mathrm{\rho} = 50$~Mm, coronal structures, we saw that $0.1$~Hz Alfv\'en waves could be fully dissipated within one density scale height which is half the height that the original \citet{rnr98} analytical solution predicts.

In Sect.~5.3 we found that, the corrected analytical solution, Eqs.~(\ref{pdsgen2})--(\ref{pdsgen1}), can accurately describe highly divergent coronal structures, even when the thin structure approximation, that the ratio of the characteristic scales in $x$ and $z$-directions is small ($x_0/H \ll 1$), used in its derivation begins to be violated. Importantly we found that $0.1$~Hz Alfv\'en waves propagating in strongly divergent, $H_\mathrm{b} = 5$~Mm, stratified coronal structures can be fully dissipated within $20$~Mm, which is again half the height that the \citet{rnr98} analytical solutions predict. This is also over six times lower than would occur as a result of standard \citet{hp83} phase mixing in uniform magnetic fields and is less than half of the density scale height. This result suggests that the importance of enhanced phase mixing as a mechanism for dissipating Alfv\'en waves in the solar corona (a stratified and divergent medium), has been seriously underestimated.

In Sect.~5.4 we investigated the viscous heating that results from enhanced phase mixing. We found that $0.1$~Hz Alfv\'en waves of amplitude $V_\mathrm{y} \approx 55$~km~s$^{-1}$ propagating in strongly divergent, $H_\mathrm{b} = 5$~Mm, stratified coronal structures, can generate viscous heating fluxes $F_\mathrm{H} \approx 2100$~J~m$^{-2}$~s$^{-1}$, with the associated viscous heating power peaking at $E_\mathrm{H} \approx 4.4 \times 10^{-4}$~J~m$^{-3}$~s$^{-1}$. This compares well to an active region's $F_\mathrm{H} \approx 100$~--~$ 2000$~J~m$^{-2}$~s$^{-1}$ and $E_\mathrm{H} \approx 4.0 \times 10^{-5}$~J~m$^{-3}$~s$^{-1}$ heating requirement \citep[see][]{a04}. Crucially, we also found that, as well as increasing the magnitude of the heating power by a factor of ten, enhanced phase mixing lowers the heating length scale by a factor of six, when compared to standard phase mixing in a uniform magnetic field. This is perhaps the most important effect, as it enables the Alfv\'en waves to dissipate their energy at low heights in the corona, within a density scale height. We therefore conclude that, given strongly divergent coronal structures, the enhanced phase mixing mechanism is a viable method for heating the corona to temperatures in excess of $1$~MK.

Finally in Sect.~5.5, we considered the dependence of the heating length scale, $L_\mathrm{H}$, on the driving frequency, $\omega$, and shear viscosity, $\nu$. Overall we found that the enhanced phase mixing of Alfv\'en waves in strongly divergent magnetic fields, can dissipate the wave energy within a density scale height, using a shear viscosity up to eight orders of magnitude lower than required by standard phase mixing in a uniform magnetic field. Specifically, we found that for strongly divergent magnetic fields, $H_\mathrm{b} = 5$~Mm, the enhanced phase mixing of observable $\omega = 0.01$~rads~s$^{-1}$ Alfv\'en waves, results in a heating length scale of $L_\mathrm{H} \le 50$~Mm, using classical Braginskii viscosity; we therefore do not need to invoke anomalous viscosity to dissipate observable Alfv\'en waves in the corona. Therefore we conclude that the value of shear viscosity required for the enhanced phase mixing mechanism to be a viable method for coronal heating, strongly depends on the heating length scale of an active region.

Our decision to focus on active regions was led by the claim by \citet{awtp07} that $83\%$ of the coronal heating requirement is located there. While our focus on strongly divergent coronal structures was necessary to meet the requirement that the heating length scale be less than the active region density scale height, $L_\mathrm{H} \le H_\mathrm{\rho}$. Given this, the potential of the enhanced phase mixing mechanism to generate significant heating to fulfill the coronal heating requirement is dependent on three critical parameters: the wave amplitude, wave frequency and the value of shear viscosity. Our use of wave amplitude $V_\mathrm{y} \approx 55$~km~s$^{-1}$ is strongly supported by observations \citep{m01,btdw98,dbp98}. Our decision to use an anomalous kinematic viscosity value of $\nu = 5 \times10^7$~m$^2$~s$^{-1}$, which is four orders of magnitude higher than that given by the \citet{b65} shear viscosity tensor, is supported by numerical evidence, \citet{t06b}, which suggests that this is indeed the effective value in the corona. We also chose to use a relatively high Alfv\'en frequency, $f = 0.1$~Hz. Given the turbulent nature of the photosphere it would seem likely that high frequency Alfv\'en waves, $f \ge 0.1$~Hz, do indeed exist; it has been suggested they are generated by micro-flaring at the chromospheric level \citep[see][]{am92}. As to whether their power spectrum is sufficient to account for coronal heating is an open question. Estimations of the high frequency Alfv\'en wave spectrum can be made by projecting back from the observed spectrum at $0.3$~AU \citep[see][]{cfk99}. Observations of high frequency Alfv\'en waves are extremely difficult, as the lack of in-situ measurements means it is not possible to measure directly the magnetic field component. Indeed in Sects.~5.2 and 5.3, we plotted the variation of the $V_\mathrm{y}$ with height, since it is the companion $B_\mathrm{y}$ component that is actually measured, via doppler line broadening. Therefore observational studies of the variation of the Alfv\'en wave $B_\mathrm{y}$ component with height should, with sufficient temporal and spatial resolution, be able to confirm or refute our model of coronal heating via the enhanced phase mixing mechanism. The upcoming ROSA (Rapid Oscillations in the Solar Atmosphere) ground based instrument, will make simultaneous observations of the photosphere, chromosphere, transition region and corona at high cadence, and may therefore be able to directly observe the propagation and dissipation of these Alfv\'en waves.

In this study, we have shown that the enhanced phase mixing mechanism can fulfill the corona heating requirement in a sufficiently divergent active region. In reaching this conclusion we note that our numerical model is relatively simple. Therefore the inclusion of additional physical effects which are known to enhance the wave dissipation, such as pressure, non-linearality and 3-dimensionality, would result in further heat deposition from the driven Alfv\'en waves. The heating results shown in this paper should therefore be considered as a lower limit. It is therefore essential to conduct realistic numerical simulations of wave dissipation using observational, 3D magnetic fields with waves driven by observed doppler shift data. Indeed, simulations incorporating observed magnetic fields have already been attempted by \citet{o07}, in relation to coronal seismology. The inclusion of physical realism into numerical simulations is crucial to fully understanding coronal heating, and must therefore form the basis of future work.

\begin{acknowledgements}
Authors acknowledge use of (a) the E. Copson math cluster funded by PPARC and the University of St. Andrews and (b) the Salford HPC Sun cluster. This work is supported by the Science and Technology Facilities Council of the United Kingdom.
\end{acknowledgements}

\begin{appendix}

\section{General solution}

Here we derive our corrected general solution, Eqs.~(\ref{pdsgen2})--(\ref{pdsgen1}), for the enhanced phase mixing of Alfv\'en waves in weakly divergent stratified coronal structures. As the derivation of the \citet{rnr98} general solution is valid up to their Eq.~(36), we will not present the full derivation here. For clarity we begin at Eq.~(33) of their analytical paper,
\begin{eqnarray}
\frac{{\partial x}}{{\partial \phi }} =  - \frac{1}{J}\frac{{\partial \psi }}{{\partial z}}, && \frac{{\partial z}}{{\partial \phi }} = \frac{1}{J}\frac{{\partial \psi }}{{\partial x}}.
\label{apn_a1}
\end{eqnarray}
Now given that $B_0 \equiv |\vec{B}_0|$ and assuming $B_0 \approx B_\mathrm{0z}  = B_{00} \partial \psi / \partial x$ gives
\begin{equation}
B_0  \approx B_{00} J\frac{{\partial z}}{{\partial \phi }},
\label{apn_a2}
\end{equation}
and therefore
\begin{eqnarray}
J \approx \frac{{B_0}}{{B_{00} }}\frac{{\partial \phi }}{{\partial z}}, && h_\phi \approx \frac{{B_0}}{{B_{00} }}\left( {\frac{{\partial z }}{{\partial \phi}}} \right)^{- 1}.
\label{apn_a3}
\end{eqnarray}
Now substitute $J$ and $\sigma = \rho_0 / \rho_{00}$ into Eq.~(26) of \citet{rnr98}
\begin{equation}
\Theta  = \int\limits_0^\Phi  {\frac{{\omega \sqrt \sigma  }}{{V_\mathrm{A0} J}}} d\Phi,
\label{apn_a4}
\end{equation}
to give
\begin{equation}
\Theta \approx \omega \varepsilon \int\limits_0^\phi  {\frac{{\sqrt{\mu_0 \rho_0 } }}{{B_0}}} \frac{{\partial z}}{{\partial \phi }}d\phi,
\label{apn_a5}
\end{equation}
We now make another approximation that $\partial \phi /\partial z \gg \partial \phi /\partial x$. The total derivative therefore becomes
\begin{equation}
d\phi  = \left( \frac{\partial \phi}{\partial x} \right)dx + \left( \frac{\partial \phi}{\partial z} \right)dz \approx \frac{{\partial \phi }}{{\partial z}}dz,
\label{apn_a6}
\end{equation}
which can then be substituted into Eq.~(\ref{apn_a5})
\begin{equation}
\Theta  \approx \omega \varepsilon \int\limits_0^z \frac{1}{V_\mathrm{A} \left( {x,z'} \right)} dz'.
\label{apn_a7}
\end{equation}
Now substitute this along with Eq.~(\ref{apn_a3}) into Eq.~(30) of \citet{rnr98}
\begin{equation}
\lambda \left( {\Phi ,\psi } \right) = \frac{\nu J h_\phi^2  \sqrt{\sigma}}{2 \varepsilon^3 V_\mathrm{A0}} \left( \frac{\partial \Theta}{\partial \psi} \right)^2,
\label{apn_a8}
\end{equation}
to give
\begin{equation}
\lambda \left( {\Phi ,\psi } \right) \approx \frac{\nu}{2\varepsilon^3 V_\mathrm{A}} \frac{B_0^2}{B_{00}^2} \frac{\partial z}{\partial \phi} \left( \frac{\partial \Theta}{\partial \psi} \right)^2,
\label{apn_a9}
\end{equation}
where from Eq.~(\ref{apn_a7})
\begin{equation}
\frac{{\partial \Theta }}{{\partial \psi }} \approx \omega \varepsilon \int\limits_0^z {\frac{\partial }{{\partial \psi }}\left( {\frac{1}{{V_\mathrm{A} \left( {x,z'} \right)}}} \right)} dz',
\label{apn_a10}
\end{equation}
Now since
\begin{equation}
\frac{\partial }{{\partial \psi }} = \frac{{\partial x}}{{\partial \psi }}\frac{{\partial V_\mathrm{A} }}{{\partial x}}\frac{\partial }{{\partial V_\mathrm{A} }},
\label{apn_a11}
\end{equation}
Eq.~(\ref{apn_a10}) becomes
\begin{equation}
\frac{{\partial \Theta }}{{\partial \psi }} \approx  - \omega \varepsilon \int\limits_0^z {\frac{1}{{V_\mathrm{A}^2 }}\frac{{\partial x}}{{\partial \psi }}\frac{{\partial V_\mathrm{A} }}{{\partial x}}} dz'.
\label{apn_a12}
\end{equation}
Next we substitute this into Eq.~(\ref{apn_a9}) to give
\begin{equation}
\lambda \left( {\Phi,\psi} \right) \approx \frac{{\omega^2 \nu }}{{2\varepsilon B_{00}^2 }}\frac{{B_0^2}}{{V_\mathrm{A}}} \left( {\int\limits_0^z {\frac{1}{{V_\mathrm{A}^2}}\frac{{\partial x}}{{\partial \psi}}\frac{{\partial V_\mathrm{A} }}{{\partial x}}} dz'} \right)^2 \frac{\partial z}{\partial \phi}.
\label{apn_a13}
\end{equation}
Now since $\Lambda \equiv \int\limits_0^\Phi  {\lambda \left( {\acute{\Phi},\psi } \right)} d\acute{\Phi}$ and using Eq.~(\ref{apn_a6}) once more, we arrive at
\begin{equation}
\Lambda(x,z) \equiv \frac{\omega^2 \nu}{2B_{00}^2} \int\limits_0^z \frac{B_0^2}{V_\mathrm{A}} \left( \int\limits_0^z \frac{1}{V_\mathrm{A}^2} \frac{\partial x}{\partial \psi} \frac{\partial V_\mathrm{A}}{\partial x} dz' \right)^2 dz'.
\label{apn_a14}
\end{equation}
which is equivalent to our general solution, Eqs.~(\ref{pdsgen2})--(\ref{pdsgen1}).

\section{Equilibrium solutions}

The error in the derivation of the \citet{rnr98} general solution also carries through into their equilibrium analytical solutions. We therefore present here derivations for these corrected equilibrium solutions. In each case we begin from our general solution for the enhanced phase mixing of Alfv\'en waves in weakly divergent stratified coronal structures, Eqs.~(\ref{pdsgen2})--(\ref{pdsgen1}).

\subsection{Non-divergent, non-stratified ($H_\mathrm{b} = H_\mathrm{\rho} = \infty$)}

Here we derive Eq.~(\ref{hp_l}); the solution first derived by \citet{hp83} for the standard phase mixing of Alfv\'en waves in non-stratified coronal structures permeated by uniform magnetic fields. From Eqs.~(\ref{flux_functions})--(\ref{densityz}) we see that
\begin{eqnarray}
B_0 = B_{00}, & & \rho_0 \left({x}\right) = \hat\rho_0 \left(x\right).
\label{apn_cs0_1}
\end{eqnarray}
Substitute these equations into Eq.~(\ref{va}) to give
\begin{equation}
V_\mathrm{A}(x) = \frac{{B_{00}}}{{\sqrt{\mu_0 \hat\rho_0} }}.
\label{apn_cs0_2}
\end{equation}
Now substitute this into Eqs.~(\ref{pdsgen2})--(\ref{pdsgen1}); our general solution
\begin{equation}
\Lambda (x,z) \approx \frac{{\omega ^2 \nu \sqrt{\mu_0 \hat\rho_0}}}{{2B_{00}}} \int\limits_0^z I^2(x,z') dz',
\label{apn_cs0_3}
\end{equation}
where the integral $I$ is given by
\begin{eqnarray}
I(x,z') =  - \frac{{\sqrt{\mu_0}}}{{2B_{00} \sqrt{\hat\rho_0}}} \frac{{\partial \hat\rho_0}}{{\partial x}} z'.
\label{apn_cs0_4}
\end{eqnarray}
Substitute this into Eq.~(\ref{apn_cs0_3})
\begin{equation}
\Lambda (x,z) \approx \frac{{\omega ^2 \nu \sqrt {\mu_0^3}}}{{8B_{00}^3 \sqrt{\hat\rho_0} }} \left({\frac{{\partial \hat\rho_0}} {{\partial x}}} \right)^2 \int\limits_0^z {z'^2} dz',
\label{apn_cs0_5}
\end{equation}
and then simplify to give
\begin{equation}
\Lambda (x,z) \approx \frac{{\omega ^2 \nu \sqrt {\mu_0^3}}}{{24B_{00}^3 \sqrt{\hat\rho_0} }} \left({\frac{{\partial \hat\rho_0}} {{\partial x}}} \right)^2 z^3.
\label{apn_cs0_6}
\end{equation}
Since $B_0 \equiv B_{00}$, this can be given in terms of $V_\mathrm{A}$ using Eq.~(\ref{apn_cs0_2})
\begin{equation}
\Lambda (x,z) \approx \frac{{\omega ^2 \nu}}{{6 V_\mathrm{A}^5}} \left({\frac{{\partial V_\mathrm{A}}} {{\partial x}}} \right)^2 z^3.
\label{apn_cs0_7}
\end{equation}
which is equivalent to Eq.~(\ref{hp_l}).

\subsection{Divergent, non-stratified ($H_\mathrm{b} \ne H_\mathrm{\rho} = \infty$)}

Now we derive Eqs.~(\ref{pds2_l})--(\ref{pds2_lb}); the solution for the enhanced phase mixing of Alfv\'en waves in non-stratified coronal structures permeated by weakly divergent magnetic fields. From Eqs.~(\ref{flux_functions})--(\ref{densityz}) we see that
\begin{eqnarray}
B_0\left(x,z\right) = B_{00} e^{ - z/H_\mathrm{b} }, & & \rho _0 \left( {x} \right) = \hat\rho_0 \left( x \right).
\label{apn_cs2_1}
\end{eqnarray}
Substitute these equations into Eq.~(\ref{va}) to give
\begin{equation}
V_\mathrm{A}(x,z) = \frac{{B_{00}}}{{\sqrt {\mu_0 \hat\rho_0 } }}e^{- z/H_\mathrm{b}}.
\label{apn_cs2_2}
\end{equation}
Now substitute this into Eqs.~(\ref{pdsgen2})--(\ref{pdsgen1})
\begin{equation}
\Lambda (x,z) \approx \frac{{\omega ^2 \nu \sqrt{\mu_0 \hat\rho_0}}}{{2B_{00}}}\int\limits_0^z {e^{ -z'/H_\mathrm{b}} I^2(x,z') dz'},
\label{apn_cs2_3}
\end{equation}
where the integral $I$ is given by
\begin{eqnarray}
I(x,z') =  - \frac{{H_\mathrm{b} \sqrt{\mu_0}}}{{4B_{00} \sqrt{\hat\rho_0}}} \sec \left(\frac{x}{H_\mathrm{b}}\right) \frac{{\partial \hat\rho_0}}{{\partial x}}\left( {e^{2z'/H_\mathrm{b}} - 1} \right).
\label{apn_cs2_4}
\end{eqnarray}
Substitute this into Eq.~(\ref{apn_cs2_3})
\begin{eqnarray}
\Lambda (x,z) & \approx & \frac{\omega^2 \nu H_\mathrm{b}^2 \sqrt {\mu_0^3}}{32B_{00}^3 \sqrt{\hat\rho_0}} \sec^2 \left(\frac{x}{H_\mathrm{b}} \right) \left( \frac{\partial \hat\rho_0}{\partial x} \right)^2 \nonumber \\
& & \int\limits_0^z {\left( e^{3z'/H_\mathrm{b}}  - 2e^{z'/H_\mathrm{b}} + e^{-z'/H_\mathrm{b}} \right)dz'}.
\label{apn_cs2_5}
\end{eqnarray}
and then simplify to give
\begin{eqnarray}
\Lambda(x,z) & \approx & \frac{ \nu \omega^2 H_\mathrm{b}^3}{96 V_\mathrm{A0}^3 \hat \rho_0^{1/2} \rho_{00}^{3/2} } \sec^2 \left( \frac{x}{H_\mathrm{b}} \right) \left( \frac{\partial \hat \rho_0}{\partial x} \right)^2 \nonumber \\
& & \left( 1 + 3e^{- z/H_\mathrm{b}} \right) \left( e^{z/H_\mathrm{b}} - 1 \right)^3.
\label{apn_cs2_6}
\end{eqnarray}
which is equivalent to Eqs.~(\ref{pds2_l})--(\ref{pds2_lb}).

\subsection{Divergent, stratified, ($H_\mathrm{b} = \frac{1}{2} H_\mathrm{\rho}$)}

Next we derive Eqs.~(\ref{pds3_l})--(\ref{pds3_lb}); the solution for the enhanced phase mixing of Alfv\'en waves in stratified coronal structures permeated by weakly divergent coronal magnetic fields where $H_\mathrm{b} = \frac{1}{2} H_\mathrm{\rho}$. From Eqs.~(\ref{flux_functions})--(\ref{densityz}) we see that
\begin{eqnarray}
B_0\left(x,z\right) = B_{00} e^{ - z/H_\mathrm{b} }, & & \rho _0 \left({x,z}\right) = \hat \rho_0 \left(x\right)e^{-2z/H_\mathrm{b}}.
\label{apn_cs3_1}
\end{eqnarray}
Substitute these equations into Eq.~(\ref{va}) to give
\begin{equation}
V_\mathrm{A}(x) = \frac{{B_{00}}}{{\sqrt{\mu_0 \hat \rho_0} }}.
\label{apn_cs3_2}
\end{equation}
Note that this is a function of $x$ only. Now substitute this into our general solution given by Eqs.~(\ref{pdsgen2})--(\ref{pdsgen1})
\begin{equation}
\Lambda (x,z) \approx \frac{{\omega ^2 \nu \sqrt{\mu_0 \hat \rho_0}}}{{2B_{00} }}\int\limits_0^z {e^{ - 2z'/H_\mathrm{b} } I^2(x,z') dz'},
\label{apn_cs3_3}
\end{equation}
where the integral $I$ is given by
\begin{eqnarray}
I(x,z') =  - \frac{{H_\mathrm{b} \sqrt{\mu_0}}}{{2B_{00} \sqrt {\hat \rho_0 } }} \sec \left( \frac{x}{H_\mathrm{b}}\right) \frac{{\partial \hat \rho_0}}{{\partial x}}\left( {e^{z'/H_\mathrm{b} }  - 1} \right).
\label{apn_cs3_4}
\end{eqnarray}
Substitute this into Eq.~(\ref{apn_cs3_3})
\begin{eqnarray}
\Lambda (x,z) & \approx & \frac{{\omega ^2 \nu H_\mathrm{b} ^2 \sqrt {\mu_0 } }}{{8B_{00} \sqrt{\hat\rho_0^{3}} }} \sec^2 \left(\frac{x}{H_\mathrm{b}} \right) \left( {\frac{{\partial \hat \rho_0 }}{{\partial x}}} \right)^2 \nonumber \\
& & \int\limits_0^z {\left( {1 - 2e^{ - z'/H_\mathrm{b} }  + e^{ - 2z'/H_\mathrm{b} } } \right)dz'},
\label{apn_cs3_5}
\end{eqnarray}
and then simplify to give
\begin{eqnarray}
\Lambda (x,z) & \approx & \frac{\omega^2 \nu H_\mathrm{b}^3}{16V_\mathrm{A} \hat \rho_0^2} \sec^2 \left( \frac{x}{H_\mathrm{b}} \right)\left( \frac{\partial \hat \rho_0}{\partial x} \right)^2 \nonumber \\
& & \left[ \left( e^{-z/H_\mathrm{b}  - 3} \right)\left( 1 - e^{- z/H_\mathrm{b}} \right) + \frac{2z}{H_\mathrm{b}} \right].
\label{apn_cs3_6}
\end{eqnarray}
which is equivalent to Eqs.~(\ref{pds3_l})--(\ref{pds3_lb}).

\subsection{Divergent, stratified ($H_\mathrm{b} \ne \frac{1}{2} H_\mathrm{\rho}$)}

Finally we derive Eqs.~(\ref{pds4_l})--(\ref{pds4_lb}); the solution for the enhanced phase mixing of Alfv\'en waves in stratified coronal structures permeated by weakly divergent magnetic fields where $H_\mathrm{b} \ne \frac{1}{2} H_\mathrm{\rho}$. From Eqs.~(\ref{flux_functions})--(\ref{densityz}) we see that
\begin{eqnarray}
B_0\left(x,z\right) = B_{00} e^{ - z/H_\mathrm{b} }, & & \rho _0 \left({x,z}\right) = \hat \rho _0 \left(x\right)e^{ - z/H_\mathrm{\rho}}.
\label{apn_cs4_1}
\end{eqnarray}
Substitute these equations into Eq.~(\ref{va}) to give
\begin{equation}
V_\mathrm{A}(x,z)  = \frac{B_{00}}{\sqrt{\mu_0 \hat \rho_0}} e^{z\left( 1/2H_\mathrm{\rho} - 1/H_\mathrm{b} \right)}.
\label{apn_cs4_2}
\end{equation}
Now substitute this into Eqs.~(\ref{pdsgen2})--(\ref{pdsgen1})
\begin{equation}
\Lambda(x,z)  \approx \frac{{\omega ^2 \nu \sqrt {\mu_0 \hat \rho_0 } }}{{2B_{00} }}\int\limits_0^z {e^{ - z'\left( {1/2H_\mathrm{\rho}   + 1/H_\mathrm{b} } \right)} I^2(x,z') dz'},
\label{apn_cs4_3}
\end{equation}
where the integral $I$ is given by
\begin{eqnarray}
I(x,z') = \frac{{\sqrt {\mu _0 } }}{{B_{00}^{} \sqrt {\hat \rho_0 } }} {\frac{{H_\mathrm{\rho} H_\mathrm{b} }}{{H_\mathrm{b}  - 4H_\mathrm{\rho} }}}  \sec \left(\frac{x}{H_\mathrm{b}} \right) \frac{\partial \hat \rho_0}{\partial x} \left( {e^{ - z'\left( 1/2H_\mathrm{\rho}  - 2/H_\mathrm{b} \right)}  - 1} \right). \nonumber
\end{eqnarray}
\begin{equation}
\label{apn_cs4_4}
\end{equation}
Substitute this into Eq.~(\ref{apn_cs4_3})
\begin{eqnarray}
\Lambda(x,z) & \approx & \frac{{\omega^2 \nu \sqrt{\mu_0^{3}} }}{{2B_{00}^3 \sqrt{\hat\rho_0} }}\left( {\frac{{H_\mathrm{\rho}  H_\mathrm{b} }}{{H_\mathrm{b}  - 4H_\mathrm{\rho}  }}} \right)^2 \sec^2 \left(  \frac{x}{H_\mathrm{b}} \right) \left( {\frac{{\partial \hat \rho_0 }}{{\partial x}}} \right)^2  \nonumber \\
& & \int\limits_0^z e^{ - z'\left( 1/2H_\mathrm{\rho} + 1/H_\mathrm{b} \right)} \left( {e^{ - z'\left( 1/2H_\mathrm{\rho} - 2/H_\mathrm{b} \right)}  - 1} \right)^2 dz',
\label{apn_cs4_5}
\end{eqnarray}
and then simplify to give
\begin{eqnarray}
\Lambda (x,z) & \approx & \frac{\nu \omega^2}{V_\mathrm{A0}^3 \hat \rho_0^{1/2} \rho_{00}^{3/2}} \frac{H_\rho^3 H_\mathrm{b}^3}{\left(4H_\rho - H_\mathrm{b} \right)^2} \sec^2 \left( \frac{x}{H_\mathrm{b}}\right) \left( \frac{\partial \hat \rho_0}{\partial x}\right)^2 \nonumber \\
& & \left[ \frac{1 - \exp \left( 3z/H_\mathrm{b} - 3z/2H_\rho \right)}{3\left( H_\mathrm{b} - 2H_\rho \right)} \right. \nonumber \\
& & + \frac{1 - \exp \left( z/H_\mathrm{b} - z/H_\rho \right)}{H_\rho - H_\mathrm{b}} \nonumber \\
& & \left. + \frac{1 - \exp \left( - z/2H_\rho - z/H_\mathrm{b} \right)}{H_\mathrm{b} + 2H_\rho} \right].
\label{apn_cs4_6}
\end{eqnarray}
which is equivalent to Eqs.~(\ref{pds4_l})--(\ref{pds4_lb}).

\end{appendix}

\bibliography{pds001}

\end{document}